\documentclass[journal]{IEEEtran}

\usepackage[final]{changes} 

\usepackage{amsmath,amsfonts}
\usepackage{algorithmic}
\usepackage{array} 
\usepackage{textcomp}
\usepackage{stfloats}
\usepackage{url}
\usepackage{verbatim}
\usepackage{graphicx}
\def\BibTeX{{\rm B\kern-.05em{\sc i\kern-.025em b}\kern-.08em
    T\kern-.1667em\lower.7ex\hbox{E}\kern-.125emX}}
\usepackage{balance}

\usepackage{cite}
\usepackage{booktabs} 
\usepackage{makecell}
\usepackage{tabularx}
\usepackage{threeparttable}
\usepackage{caption}
\usepackage{multirow}

\usepackage{array}

\newcolumntype{C}{>{\centering\arraybackslash}X}
\newcolumntype{L}{>{\raggedleft\arraybackslash}X}
\newcolumntype{R}{>{\raggedright\arraybackslash}X}

\usepackage{titlesec}
\titlespacing*{\section}
{0pt}{0.8ex plus 0.3ex minus 0.2ex}{0.6ex}
\titlespacing*{\subsection}
{0pt}{0.6ex plus 0.2ex}{0.4ex}
\titlespacing*{\subsubsection}
{0pt}{0.4ex}{0.3ex}

\newcommand{\secref}[1]{Section~\ref{#1}}
\newcommand{\fgref}[1]{Fig.~\ref{#1}}
\newcommand{\tabref}[1]{Table~\ref{#1}}

\usepackage{xcolor}
\newcommand{\revision}[1]{#1}

\usepackage{subfigure}
\usepackage{subcaption}
\usepackage[ruled, linesnumbered, vlined]{algorithm2e}  
\usepackage[colorlinks,linkcolor=red,anchorcolor=blue,citecolor=green]{hyperref}

\begin{document}
\title{
SparseForge: Beyond Spike Sparsity toward Hardware-Exploitable Neuron Pruning for Efficient SNN Acceleration
}
\author{
Kainan Wang, Sanhe Fu and Aili Wang,~\IEEEmembership{Member,~IEEE} 
\thanks{
This work was supported in part by the National Natural Science Foundation of China under grant 62304203, in part by the International Campus of ZJU International Research Collaboration Seed Project under grant GJ202509, and in part by the  International Campus of ZJU startup fund. 
\textit{(Corresponding author: Aili Wang.)} 
}
\thanks{Kainan Wang, Sanhe Fu and Aili Wang are with the Zhejiang University-University of Illinois Urbana-Champaign Institute, Zhejiang University, Haining, Zhejiang China. (e-mail: \{kainan.22, sanhe1.25, ailiwang\}@intl.zju.edu.cn).}
}

\maketitle

\begin{abstract}  
Spiking Neural Networks (SNNs) provide an event-driven paradigm for energy-efficient edge computing. However, operation-level sparsity in SNNs does not necessarily eliminate the hardware costs associated with redundant output neurons, including neuron-state maintenance, partial-sum (psum) accesses, and on-chip storage requirements. This paper presents SparseForge, a hardware-aware sparse SNN acceleration framework that jointly
optimizes neuron pruning, sparse dataflow, and accelerator microarchitecture.
At the algorithm level, we introduce a gated leaky integrate-and-fire (G-LIF) neuron to learn differentiable neuron masks with binary spike semantics. A dual regularization scheme is further developed to align pruning with both synaptic-operation reduction and psum buffer capacity reduction.
At the dataflow level, we design a sparse Column-Wise Product (CWP) dataflow with a variant of incremental spike encoding to improve weight-access regularity and reduce
sparse-index overhead under dynamic spike activity. At the architecture level, we develop a mask-aware sparse processing unit (SPU) that supports compressed spike
decoding, sparse implicit im2col conversion, neuron-mask-gated psum updates, and decoupled sparse execution.
Experimental results across multiple SNN backbones and datasets show that SparseForge retains 1.75\%--77.99\% of neurons and 0.16\%--62.36\% of synaptic operations, with a configurable accuracy--efficiency trade-off. The buffer-capacity-aware regularization reduces the required psum buffer capacity by up to 96.18\%. FPGA implementations on a Xilinx ZCU9EG achieve 296.94--619.49 GOPS/W across the evaluated workloads. 
\revision{
Across matched network and dataset settings, SparseForge provides $1.24\times$--$4.17\times$ higher FPS and $1.50\times$--$7.18\times$ higher GOPS/W than representative FPGA-based SNN accelerators, while reducing EDP by $1.70\times$--$19.13\times$, with benchmark-dependent accuracy--efficiency trade-offs.
}

\end{abstract}

\begin{figure*}[tbp]
    \centering 
    \vspace{-8pt}
    \includegraphics[width=0.95\linewidth]{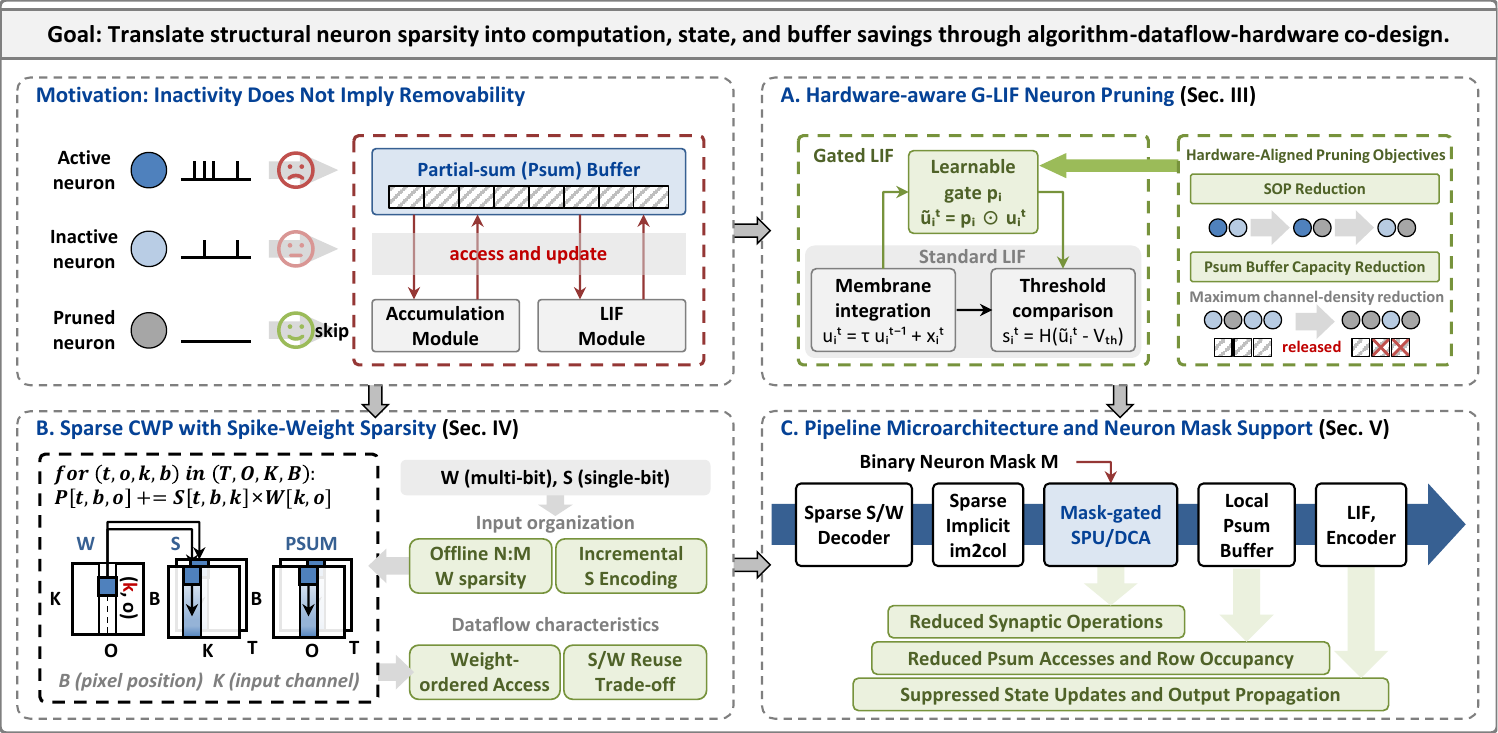}
    \caption{\revision{Overview of SparseForge. (a) Hardware-aware G-LIF pruning generates deterministic neuron masks, (b) sparse CWP with incremental spike encoding skips zero-valued operations while preserving ordered weight access, and (c) the mask-aware accelerator translates the resulting sparsity into computation and storage savings.}}
    \label{overview}
    \vspace{-15pt}
\end{figure*}

\section{Introduction}
\label{introduction}
Spiking Neural Networks (SNNs) are promising for energy-efficient edge computing because their binary and temporally sparse activations enable event-driven execution \cite{Wolfgang}. However, reducing spike-triggered operations does not necessarily eliminate the hardware costs associated with output neurons. Neurons with low activity may still require membrane-potential updates, partial-sum (psum) accesses, buffer capacity, and output paths. Therefore, translating algorithmic sparsity into reductions in both computation and storage remains an important challenge.

Existing FPGA accelerators exploit SNN sparsity at different granularities. Early designs, including DeepFire2 \cite{Deepfire2}, FireFly \cite{Firefly}, and FireFly2 \cite{Firefly2}, improve throughput through deeply pipelined datapaths and optimized DSP-slice and memory organizations. Spike-sparsity accelerators such as SpinalFlow \cite{Spinalflow}, Parallel Time Batching (PTB) \cite{PTB}, and STELLAR \cite{Stellar} further reduce event processing and data movement through compressed spike streams or temporal parallelism. More recent designs coordinate spike and weight sparsity. LoAS \cite{Loas} employs a fully temporal-parallel inner-product dataflow, whereas GustavSNN \cite{Gustavsnn} adopts a column-parallel Gustavson-product dataflow to reduce sparse-operand matching and local-state overhead. FireFly-S \cite{FireFlyS}, Cactus \cite{Cactus}, and FPSpike \cite{FPSpike} exploit dual or structured sparsity to improve computation and storage efficiency. \revision{In addition, Prosperity \cite{Prosperity}, ExSpike \cite{ExSpike}, and Phi \cite{Phi} reuse common activation patterns, event contributions, or hierarchical sparse representations. Despite these advances, these accelerators primarily optimize synaptic products and generally retain the original output-neuron space, limiting reductions in neuron-associated state and buffer resources.}

Algorithmic approaches have also been developed to reduce SNN activity and network size. Activity-oriented training suppresses spike events but does not identify neurons that can be deterministically removed \cite{BuTong}. Existing structural pruning methods mainly optimize neuron count, accuracy, or operation count without considering how the retained-neuron distribution affects physical buffer provisioning \cite{ShiXinyu}. NPTD \cite{NPTD} terminates future computation using time-dependent thresholds, but its dynamic decisions are less compatible with fully temporal-parallel and deeply pipelined execution. Moreover, it skips only the remaining computations of selected neurons rather than removing them from the mapped network.

To address these limitations, we propose SparseForge, a unified algorithm--dataflow--architecture co-design framework for hardware-exploitable structural neuron sparsity. SparseForge learns deterministic neuron masks that suppress redundant synaptic computation, state updates, psum storage, and output propagation. Its optimization objective accounts for both synaptic operation (SOP) reduction and the maximum retained-neuron density that determines psum buffer capacity. A sparse Column-Wise Product (CWP) dataflow and incremental spike encoding preserve ordered weight access while reducing sparse-index overhead, and a mask-aware accelerator translates the learned structure into hardware execution and storage savings.

Experiments across six SNN configurations show that event-based workloads provide particularly large compression opportunities. On DVS128-Gesture \cite{dvs128_Gesture}, SparseForge retains only 6.76\% of neurons and 1.51\% of SOPs, saves 88.60\% of psum buffer capacity, and improves accuracy by 0.69 percentage points. It also remains effective on frame-based workloads; on SCNN6/CIFAR-10 \cite{cifar}, it reduces neuron count, SOPs, and psum buffer capacity by 83.00\%, 92.40\%, and 74.00\%, respectively, with a 1.62-percentage-point accuracy loss. FPGA implementations on a Xilinx ZCU9EG achieve 296.94--619.49 GOPS/W and improve throughput, energy efficiency, and EDP over representative accelerators under matched benchmarks.

The main contributions are summarized as follows:
\begin{itemize}
    \item We propose a structural pruning method based on gated leaky integrate-and-fire (G-LIF) neurons, with SOP- and buffer-capacity-aware regularization that learns deterministic masks to reduce computation, psum accesses, and storage requirements.
    \item We develop a sparse CWP dataflow with incremental spike encoding to skip zero spikes and weights, preserve ordered sparse-weight access, and compactly represent dynamic spike activity.
    \item We design a mask-aware accelerator supporting compressed decoding, implicit im2col conversion, gated psum updates, and temporally and spatially decoupled sparse execution.
\end{itemize}
The remainder of this paper is organized as follows. \secref{background} analyzes SNN inference costs and summarizes the requirements for hardware-exploitable sparsity from the perspectives of neuron masking, sparse dataflow, and microarchitecture support. \secref{three} presents the proposed G-LIF-based neuron pruning method with SOP and buffer-capacity-aware regularization. \secref{four} describes the sparse CWP dataflow and sparse spike encoding. \secref{five} details the SparseForge accelerator architecture and its mask-aware sparse processing units. \secref{six} provides experimental results and ablation studies, and \secref{seven} concludes this paper.

\section{Background and Motivation} \label{background}  
\subsection{SNN Inference Cost Beyond Spike Operations}
SNNs encode information with temporally distributed binary spike events. For the $l$-th layer, a standard leaky integrate-and-fire (LIF) neuron \cite{lif} integrates the synaptic input $x_l^t$ into the membrane potential $u_l^t$ and generates an output spike $s_l^t$ when the membrane potential exceeds the threshold:
\begin{equation}
\begin{aligned}
u_l^t
&=
\tau u_l^{t-1} + x_l^t, \\
s_l^t
&=
H(u_l^t - V_{\rm th})
\end{aligned}
\label{eq:glif_dynamics}
\end{equation}
where $\tau$ is the leakage factor, $V_{\rm th}$ is the firing threshold, and $H(\cdot)$ denotes the Heaviside function.

In the convolutional SNN workloads considered in this work, spike-triggered synaptic accumulations constitute a major arithmetic cost. For a convolutional layer, the synaptic operation count can be written as:
\begin{equation}
{\rm SOP}_{l} =
\sum_{t=1}^{T}
\sum_{j \in \mathcal{N}^{\rm out}_l}
m_{l,j}
\sum_{i \in \mathcal{R}_f(j)}
s_{l-1,i}^{t}
\end{equation}
where $T$ denotes the number of time steps,
$\mathcal{N}^{\rm out}_l$ denotes the index set of output neurons in
layer $l$, $m_{l,j}\in\{0,1\}$ indicates whether output neuron $j$ is
retained, and $\mathcal{R}_f(j)$ denotes the receptive field of neuron
$j$. This expression shows that the synaptic cost depends jointly on
input spike activity and retained output-neuron paths. 
Although reducing spike-triggered operations can substantially improve computational efficiency, operation-level sparsity alone does not necessarily eliminate the hardware resources associated with output neurons. Neurons with low firing rates may still incur state-storage, state-update, psum buffer, and control overheads, as illustrated in \fgref{overview}. This observation motivates structural neuron pruning with explicit inference-time masks to disable redundant neuron-associated computation, state updates, and output propagation.

\subsection{Hardware-Exploitable Structural Neuron Sparsity}
Unlike dynamic spike sparsity, structural neuron sparsity provides fixed retention information that can be incorporated into accelerator mapping and resource provisioning. Such hardware benefits become more accessible when the learned pruning decisions are converted into deterministic binary masks for inference.

To translate the learned sparsity into an accelerator-compatible structure, the pruning formulation is guided by three considerations. First, neuron-retention variables should remain differentiable during training to support joint optimization with the task objective. Second, the learned decisions should be hardened into fixed binary masks for deterministic inference. Third, the optimization objective should reflect both spike-triggered synaptic operations and the physical psum buffer capacity associated with the retained-neuron distribution. In the proposed accelerator, this capacity is determined by the maximum number of retained neurons among independently addressed output-channel buffers, rather than by the overall neuron-sparsity ratio. These considerations motivate the G-LIF neuron pruning and buffer-capacity-aware regularization introduced in \secref{three}.

\subsection{Sparse Dataflow Considerations}
Sparse dataflow efficiency depends not only on skipping zero-valued operations but also on the regularity of operand access and partial-sum accumulation. Encoding both operands sparsely may require runtime index matching, whereas fully event-driven traversal can make weight accesses and psum updates dependent on irregular activation patterns. The dataflow should therefore balance zero skipping, scheduling regularity, and sparse-index processing.

This balance is influenced by the asymmetric properties of SNN operands: spike activations are binary and vary dynamically across time, whereas synaptic weights remain static during inference and can be structured offline. These properties favor a weight-oriented traversal in which pre-arranged sparse weights establish the execution order, while decoded spike events determine whether the associated accumulations are performed. Based on this principle, sparse CWP prioritizes ordered weight access and event-level spike skipping, with incremental spike encoding used to reduce index-storage and decoding overhead. The resulting dataflow and encoding scheme are detailed in \secref{four}.

\section{Hardware-Aware G-LIF Neuron Pruning}  \label{three}
\subsection{Unified Hardware-Aware Optimization Objective}

We optimize neuron retention jointly with the task objective using the following hardware-aware loss:
\begin{equation}
\mathcal{L} =
\mathcal{L}_{\mathrm{task}}
+
\lambda_{\mathrm{sop}} R_{\mathrm{comp}}
+
\lambda_{\mathrm{buf}} R_{\mathrm{buf}}
\label{eq:unified_objective}
\end{equation}
where $\mathcal{L}_{\mathrm{task}}$ denotes the task loss, 
$R_{\mathrm{comp}}$ is a differentiable surrogate for spike-triggered synaptic operations, 
and $R_{\mathrm{buf}}$ characterizes the psum buffer capacity induced by the retained-neuron distribution. 
The coefficients $\lambda_{\mathrm{sop}}$ and $\lambda_{\mathrm{buf}}$ control the respective regularization strengths. 
These two terms capture distinct hardware effects: $R_{\mathrm{comp}}$ depends jointly on spike activity and neuron retention, whereas $R_{\mathrm{buf}}$ is governed by the maximum retained-neuron density among independently addressed output-channel buffers. 
Their formulations are presented in \secref{3c} and \secref{3d}, respectively.

\subsection{G-LIF Neuron with Learnable Structural Gates}
To learn neuron-retention decisions through gradient-based optimization, we introduce G-LIF neuron. Its membrane dynamics are given by
\begin{equation}
\begin{aligned}
u_l^t
&=
\tau u_l^{t-1} + x_l^t, \\
\widetilde{u}_l^t
&=
G_l \odot u_l^t, \\
s_l^t
&=
H\left(\widetilde{u}_l^t - V_{\mathrm{th}}\right),
\end{aligned}
\label{eq:glif_dynamics}
\end{equation}
where $\odot$ denotes element-wise multiplication. The structural gate is shared across all time steps, providing a time-invariant retention variable for each neuron.

For layer $l$, the continuous gate is parameterized as
\begin{equation}
G_l
=
\sigma\left(\rho(k)\alpha_l\right),
\qquad
G_l \in (0,1)^{C_l^o \times H_l^o \times W_l^o},
\label{eq:continuous_gate}
\end{equation}
where $\alpha_l$ denotes the learnable gate logits, $k$ is the global training iteration, $\rho(k)$ is a sharpening schedule, and $\sigma(\cdot)$ denotes the sigmoid function. As training progresses, $\rho(k)$ gradually drives the gate values toward binary decisions.

The corresponding binary retention mask is obtained by thresholding the continuous gate:
\begin{equation}
M_l
=
\mathbb{I}\left[G_l > \theta\right],
\label{eq:binary_mask}
\end{equation}
where $\theta$ is the binarization threshold. To optimize regularization terms defined on the thresholded retention decisions, we construct a straight-through mask as
\begin{equation}
M_l^{\mathrm{ste}}
=
\operatorname{sg}
\left(
M_l-G_l
\right)
+
G_l,
\label{eq:ste_mask}
\end{equation}
where $\operatorname{sg}(\cdot)$ denotes the stop-gradient operator. Thus, $M_l^{\mathrm{ste}}$ equals the binary mask $M_l$ in the forward pass while propagating gradients through $G_l$ during backpropagation. After training, $M_l$ is fixed and replaces $G_l$ in Eq.~\eqref{eq:glif_dynamics} for deterministic inference and accelerator mapping.

\subsection{Computation-Aware SOP Regularization}
\label{3c}
To account for the computation associated with both spike activity and neuron retention, we define the normalized SOP regularizer as
\begin{equation}
R_{\mathrm{comp}}
=
\sum_{l \in \mathcal{P}}
\frac{
\widehat{\mathrm{SOP}}_l
}{
\mathrm{SOP}_{\mathrm{ref},l}+\epsilon
},
\label{eq:sop_regularization}
\end{equation}
where $\mathcal{P}$ denotes the set of prunable layers, $\mathrm{SOP}_{\mathrm{ref},l}$ is the reference SOP count of the corresponding unpruned layer, and $\epsilon$ is a small constant for numerical stability. The layer-wise normalization prevents layers with larger feature maps or channel dimensions from dominating the regularization objective.

For layer $l$, the differentiable SOP estimate is given by
\begin{equation}
\widehat{\mathrm{SOP}}_l
=
\frac{1}{B}
\sum_{b=1}^{B}
\sum_{t=1}^{T}
\sum_{r=1}^{R_l}
\sum_{c=1}^{C_l^o}
G_l(r,c)\,
\kappa_l^{b,t}(r),
\label{eq:sop_estimate}
\end{equation}
where $B$ is the batch size, $T$ is the number of time steps, $R_l=H_l^oW_l^o$ is the number of output spatial positions, and $C_l^o$ is the number of output channels. The term $\kappa_l^{b,t}(r)$ denotes the number of input spike events within the receptive field of output position $r$ for sample $b$ at time step $t$.

Weighting $\kappa_l^{b,t}(r)$ by the soft gate $G_l(r,c)$ assigns a larger regularization cost to retaining neurons associated with more spike-triggered synaptic operations. The resulting objective therefore optimizes the computational effect of neuron retention rather than treating all retained neurons as equally costly.

\subsection{\revision{Buffer-Capacity-Aware Density Regularization}}
\label{3d}
Reducing the total number of retained neurons does not necessarily reduce the physical psum buffer capacity. In the proposed accelerator, $P_{co}$ represents the number of output channels processed in parallel, with each output-channel lane maintaining an independently addressed psum buffer. An unpruned layer requires a depth of $R_l$ for each lane. Under compressed addressing, the provisioned depth is instead determined by the largest retained-neuron count among the output channels mapped to the same buffer configuration. Controlling this maximum enables capacity reduction without requiring neuron masks to be spatially aligned across channels.

Unlike $R_{\mathrm{comp}}$, which estimates synaptic computation using the soft gate $G_l$, the buffer regularizer operates on the thresholded decisions represented by $M_l^{\mathrm{ste}}$. We first define the set of output spatial configurations as
\begin{equation}
\mathcal{S}
=
\left\{
\left(H_l^o,W_l^o\right)
\mid
l\in\mathcal{P}
\right\}.
\label{eq:spatial_configurations}
\end{equation}
For each configuration $s\in\mathcal{S}$, the corresponding layers are grouped as
\begin{equation}
\mathcal{L}_s
=
\left\{
l\in\mathcal{P}
\mid
\left(H_l^o,W_l^o\right)=s
\right\}.
\label{eq:buffer_group}
\end{equation}
All layers in $\mathcal{L}_s$ share the same number of output spatial positions, denoted by $R_s$.

For output channel $c$ in layer $l$, the post-threshold retention density is
\begin{equation}
D_{l,c}
=
\frac{1}{R_l}
\sum_{r=1}^{R_l}
M_l^{\mathrm{ste}}(r,c).
\label{eq:channel_density}
\end{equation}
In the forward pass, $M_l^{\mathrm{ste}}$ has the same value as the binary mask $M_l$; hence, $D_{l,c}$ corresponds to the density used for deployment-time capacity estimation.
The largest channel density associated with configuration $s$ is approximated as
\begin{equation}
\widetilde{D}_s =
\operatorname{SmoothMax}_{\tau_D}
\left(
\left\{
D_{l,c}
\mid
l\in\mathcal{L}_s,\;
1\leq c\leq C_l^o
\right\}
\right),
\label{eq:smooth_max_density}
\end{equation}
where the normalized smooth-maximum function is defined as
\begin{equation}
\operatorname{SmoothMax}_{\tau_D}
(x_1,\ldots,x_N)
=
\tau_D
\log
\left[
\frac{1}{N}
\sum_{i=1}^{N}
\exp\left(\frac{x_i}{\tau_D}\right)
\right].
\label{eq:normalized_smooth_max}
\end{equation}
Here, $N$ denotes the number of densities included in the aggregation, and $\tau_D$ is a fixed smoothing temperature. The normalization removes the dependence on the number of aggregated channels, while a smaller $\tau_D$ yields a closer approximation to the maximum.

The buffer-capacity-aware regularizer is defined as
\begin{equation}
R_{\mathrm{buf}}
=
\sum_{s\in\mathcal{S}}
\omega_s\widetilde{D}_s,
\label{eq:buffer_regularization}
\end{equation}
where the hardware-aware weight
\begin{equation}
\omega_s
=
\frac{
N_s^{\mathrm{SPU}}P_{co,s}R_s
}{
\displaystyle
\sum_{s'\in\mathcal{S}}
N_{s'}^{\mathrm{SPU}}P_{co,s'}R_{s'}
}
\label{eq:buffer_weight}
\end{equation}
reflects the relative physical buffer capacity associated with configuration $s$. Here, $N_s^{\mathrm{SPU}}$ denotes the number of sparse processing unit (SPU) groups containing psum buffers for configuration $s$, and $P_{co,s}$ denotes the corresponding output-channel parallelism. Minimizing $R_{\mathrm{buf}}$ aligns the learned retention patterns with the maximum per-channel density that governs physical buffer provisioning.

\begin{figure}[tbp]
    \centering 
    \includegraphics[width=\linewidth]{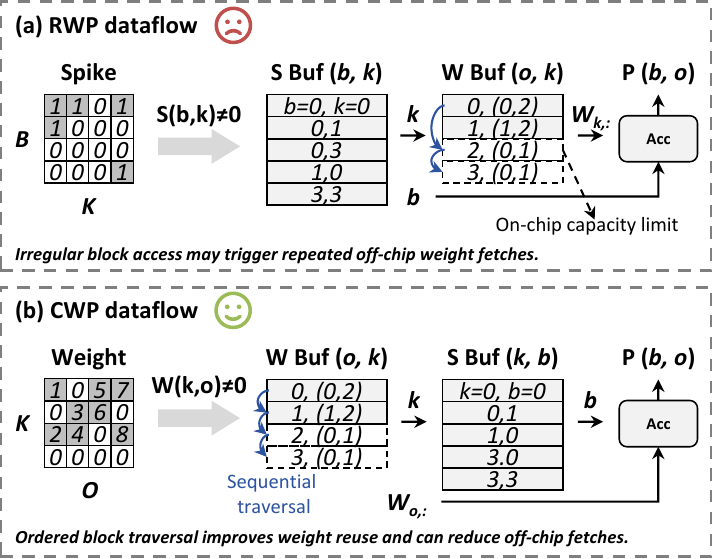}
    \caption{\revision{RWP versus sparse CWP: (a) irregular spike-driven weight access; (b) sequential weight-driven execution.}}
    \label{dataflow}
    \vspace{-15pt}
\end{figure}

\section{CWP Dataflow for Sparse Spike-Driven Inference} \label{four}  
\label{sec:cwp_dataflow}

\subsection{Sparse CWP Dataflow for Dual-Sparse Spike-Driven Inference}
\label{subsec:dataflow_design}
During SNN inference, spike activations are binary, time-varying, and dynamically sparse, whereas synaptic weights are static and can be pre-arranged offline. Therefore, an efficient sparse dataflow is expected to retain the event-driven benefits of spike sparsity while maintaining regular weight-side access.

\revision{
As shown in \fgref{dataflow} (a), a conventional row-wise product (RWP) dataflow is naturally driven by spike events. When a spike occurs at a spatial position, the accelerator accesses the corresponding weight blocks and updates the related output partial sums. Although this execution scheme directly skips zero inputs, its weight-block access order is determined by the runtime spike pattern. Consequently, successive spike events may request non-contiguous weight blocks, as indicated by the blue arrows in \fgref{dataflow} (a). When the required weight working set exceeds the capacity of the on-chip weight buffer (W-Buffer), previously evicted blocks may be requested again, potentially triggering repeated off-chip weight fetches. This problem persists under structured weight sparsity because the retained blocks are still accessed in a runtime-dependent order.
}

\revision{
In contrast, the sparse CWP dataflow shown in \fgref{dataflow} (b) adopts a weight-block-driven execution scheme. Retained weight blocks are pre-arranged offline and traversed sequentially in a predefined order. For each resident weight block, its metadata selects the corresponding spike-column group from the spike buffer (S-Buffer), whose valid events are decoded to drive event-based accumulation. Therefore, sparse CWP preserves spike-dependent computation while maintaining sequential and deterministic weight-side access, which can reduce repeated off-chip fetches under limited W-Buffer capacity. A controlled comparison with RWP is presented in Section~\ref{sec:rwp_cwp_ablation}.
}

\begin{figure}[tbp]
    \centering 
    \includegraphics[width=0.95\linewidth]{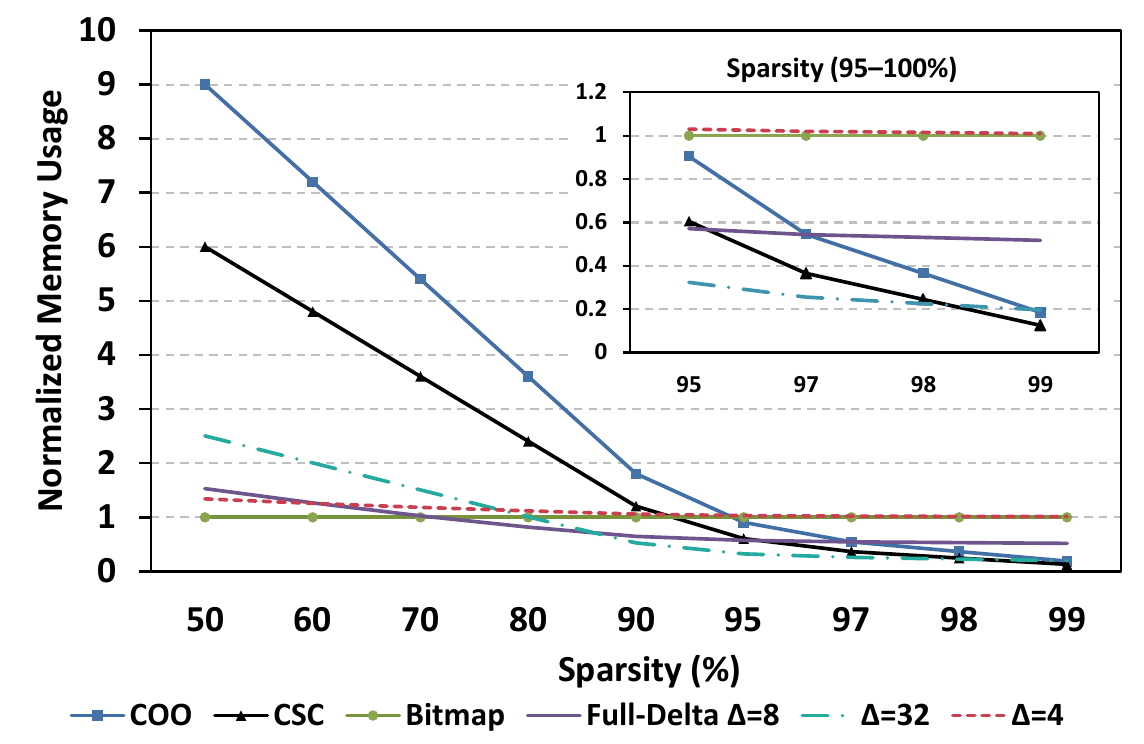}
    \vspace{-5pt}
    \caption{Normalized spike-index storage cost of Full-Delta encoding with different $\Delta$ values, compared with COO, CSC, and bitmap formats across spike sparsity levels.}
    \label{fig:activation_encoding}
    \vspace{-15pt}
\end{figure}

\subsection{Incremental Activation Representation for Sparse CWP}
\label{fourb}
In sparse CWP, the S-Buffer supplies sparse spike events from the input-column groups selected by each resident weight block. The activation representation therefore directly affects the S-Buffer capacity, on- and off-chip data-movement costs, and event-decoding overhead. Conventional coordinate (COO) and compressed sparse column (CSC) formats require explicit indices or pointers, whereas bitmap formats allocate storage for the entire spatial region regardless of spike activity. To reduce this metadata overhead while supporting ordered access, we adopt an incremental spike representation based on delta encoding.

For each $(t,b,c)$ feature map, the nonzero spatial positions are first
extracted and sorted by their linear addresses:
\begin{equation}
\mathcal{P}_{t,b,c}=\{p_1,p_2,\ldots,p_K\}, \quad
0 \le p_1 < p_2 < \cdots < p_K < H W
\end{equation}
where $p_i=h_iW+w_i$ and $K$ is the number of nonzero spike events.
Instead of storing each absolute position $p_i$, the proposed encoding
stores the gap between adjacent events:
\begin{equation}
v_i = p_i - p_{i-1} - 1,\quad p_0=-1
\end{equation}
Thus, $v_i$ represents the number of inactive spatial positions before
the $i$-th valid spike. This representation enables sequential decoding of the selected spike-column streams, naturally aligning with the weight-block-driven sparse CWP pipeline.

Each delta value is encoded using a fixed-width codeword of
$w_\Delta=\lceil \log_2 \Delta \rceil$ bits. The all-one codeword is
reserved as an extension flag, denoted as EXT, while the remaining
codewords encode residual gaps. Let $G=2^{w_\Delta}-1$ be the number of
normal residual codewords. For an arbitrary gap $v_i$, the encoder emits
$\lfloor v_i/G \rfloor$ EXT codewords followed by one residual codeword
$r_i=v_i-\lfloor v_i/G \rfloor G$. Therefore, long gaps can be
represented without increasing the codeword width.

To support regular block-wise access in the S-Buffer, the generated
delta stream is further packed into fixed-length blocks, where each
block contains $N_s$ codewords. For each $(t,b,c)$, the number of
generated codewords and the corresponding index payload are:
\begin{equation}
L_{t,b,c}=\sum_{i=1}^{K}\left(1+\left\lfloor\frac{v_i}{G}\right\rfloor\right),
\quad
M_{t,b,c}=\left\lceil\frac{L_{t,b,c}}{N_s}\right\rceil
\label{eq:delta_block_count}
\end{equation}
\begin{equation}
B_{\Delta}(t,b,c)=M_{t,b,c}N_s w_\Delta
\label{eq:delta_storage}
\end{equation}
The last block is padded when necessary. Although this introduces minor
padding overhead, it simplifies address generation, burst-like buffer
fetching, and parallel scheduling in the S-Buffer.

The decoder reconstructs valid spike positions in a streaming manner. It
maintains the current reconstructed position $\hat{p}$ and an
accumulated gap $g$. When an EXT codeword is received, $g$ is increased
by $G$. When a residual codeword $r$ is received, the next valid position
is reconstructed as:
\begin{equation}
\hat{p} \leftarrow \hat{p} + g + r + 1
\label{eq_decode}
\end{equation}
and $g$ is reset to zero. This process only requires flag comparison,
addition, and a position counter, enabling lightweight sparse-event
delivery for the SPU pipeline.

\revision{
Based on the number of packed blocks in Eq.~\eqref{eq:delta_block_count}, the storage cost of the proposed Full-Delta representation is calculated using Eq.~\eqref{eq:delta_storage}. \fgref{fig:activation_encoding} compares its normalized index-storage cost with those of COO, CSC, and bitmap representations under different spike sparsity levels. Three configurations, $\Delta \in \{4, 8, 32\}$, corresponding to codeword widths $w_{\Delta} \in \{2, 3, 5\}$ bits, are evaluated.
Bitmap provides regular access but cannot benefit from increased spike sparsity, while COO and CSC still carry non-negligible coordinate or pointer overhead. In
contrast, the proposed delta encoding reduces activation-index storage over a wide sparsity range. A smaller $\Delta$ benefits moderate sparsity due to narrower codewords, whereas a larger $\Delta$ better covers long gaps at high sparsity. In this work, $\Delta$ is treated as a configurable hardware parameter and selected according to the sparsity and gap distribution of the target layer.
}

\begin{figure}[tbp]
    \centering 
    \includegraphics[width=0.95\linewidth]{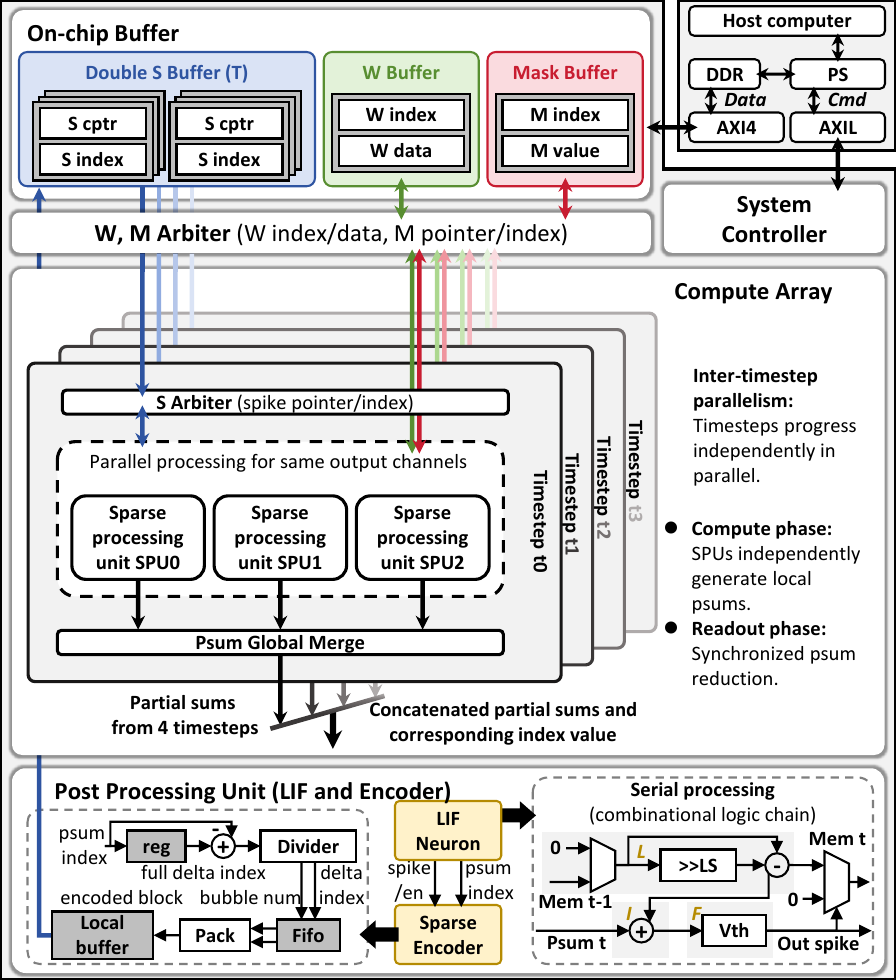}
    \caption{\revision{Overall hardware architecture of the proposed SparseForge accelerator.}}
    \label{ha}
    \vspace{-15pt}
\end{figure}

\section{Microarchitecture Design for Sparse Convolution Realization} \label{five}  
\subsection{Hardware Architecture Overview}
\fgref{ha} illustrates the proposed architecture for sparse SNN convolution. It employs a block-reuse pipeline comprising an off-chip memory interface, a system controller, on-chip buffers, a decoupled compute array (DCA), and a post-processing unit. Through the AXI interface, the host processor transfers configuration parameters, delta-encoded spike streams, structured sparse weights, and hardened neuron masks to the accelerator. The system controller decodes the commands and coordinates data movement and computation across these modules.
The on-chip storage consists of a double-buffered S-Buffer, a W-Buffer, and a mask buffer (M-Buffer). They store delta-encoded spike streams, CWP-ordered sparse weight blocks, and compact bit-vector neuron masks, respectively. The DCA integrates multiple sparse processing units, each responsible for a subset of output channels and equipped with a local psum buffer. The post-processing unit performs membrane-potential updates, threshold comparison, spike generation, and output-spike encoding. By integrating hardened neuron masks with sparse CWP execution, the architecture skips both inactive spike events and pruned neuron paths.

\subsection{Temporally and Spatially Decoupled Execution}
Under sparse CWP, convolution proceeds along the $C_{\mathrm{in}}\times K^2$ reduction dimension, covering the input-channel and kernel-spatial dimensions.
For each output neuron, synaptic contributions are progressively accumulated into a timestep-specific psum, which remains resident until all relevant weight blocks and spike events have been processed. Due to varying spike workloads, psum accumulation across timesteps may complete at different times, whereas neuron updates and spike generation must follow the temporal order of SNN dynamics.
Therefore, the proposed DCA temporally decouples psum accumulation across timesteps.
Each timestep advances according to its spike activity, avoiding lock-step stalls caused by workload imbalance. Completed psums are committed to LIF state updates in timestep order, thereby preserving temporal dependencies.

The DCA also spatially decouples execution within each output-channel group. Multiple SPUs collaboratively process the same group by partitioning the $ K^2$ kernel-spatial dimension among them. Each SPU processes its assigned kernel positions across the input channels and independently accumulates a local psum. The local psums are then reduced into the final psum before the LIF update.
The number of valid spike events may vary across kernel sub-blocks, leading to workload imbalance among SPUs. Lock-step execution would therefore stall lightly loaded SPUs. A local psum buffer in each SPU enables independent accumulation and avoids contention on a shared write-back path. This scheduling and buffering strategy improves SPU utilization while preserving the temporal semantics of SNN inference.

\begin{figure*}[tbp]
    \centering 
    \vspace{-8pt}
    \includegraphics[width=0.95\linewidth]{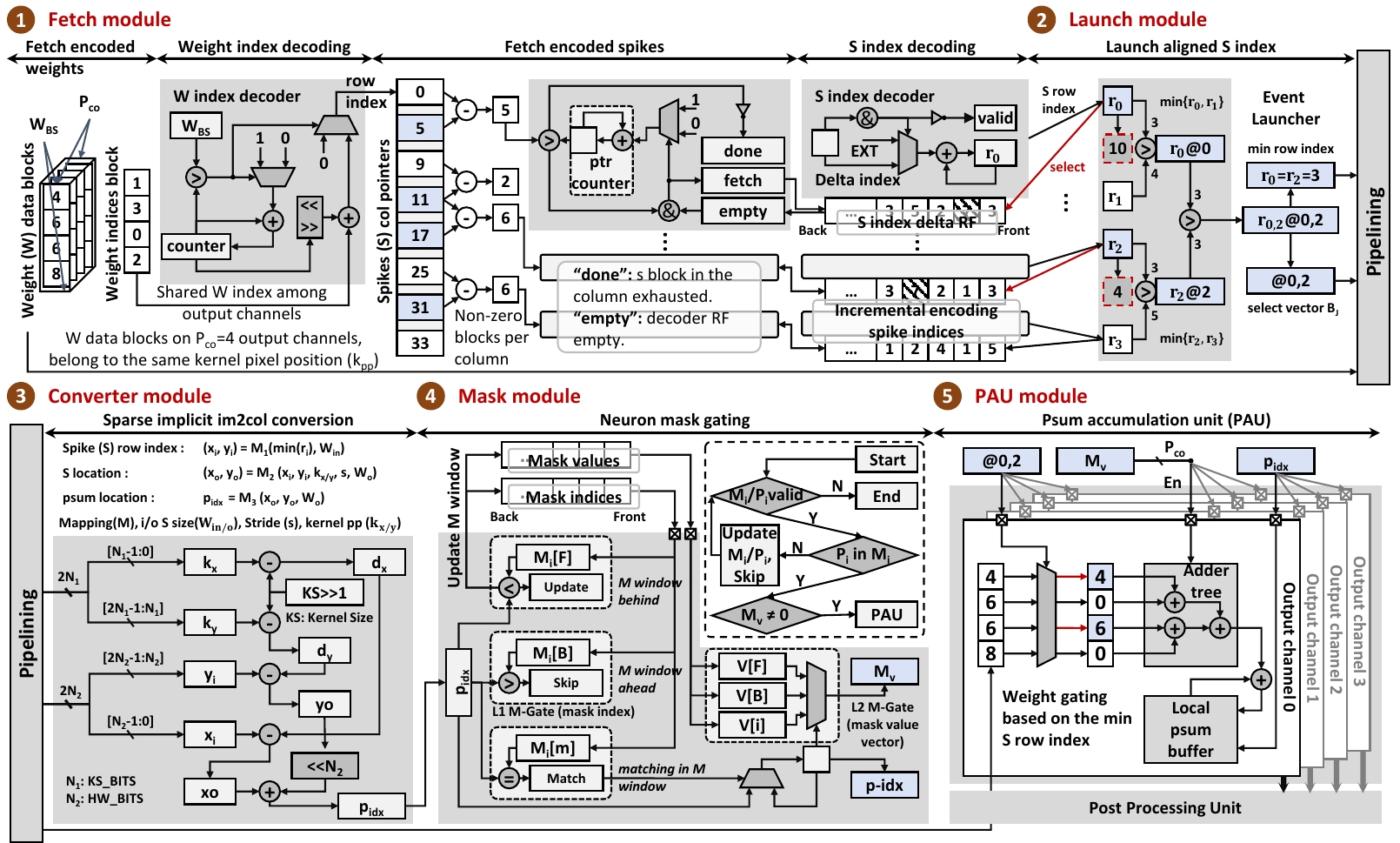}
    \caption{\revision{Microarchitecture of the SPU. The SPU consists of the fetch module, launch module, sparse implicit im2col converter, mask-gated filtering module, and weight-accumulation and psum update unit.}}
    \label{spu}
    \vspace{-15pt}
\end{figure*}

\subsection{Sparse Processing Unit Microarchitecture}
\fgref{spu} illustrates the detailed microarchitecture of the proposed SPU, which comprises five pipelined modules: fetch, launch, implicit im2col conversion, mask filtering, and psum accumulation unit (PAU). The fetch module loads compressed weights and spike streams and decodes their indices, while the launch module dispatches aligned sparse events. The implicit im2col module converts these events into convolution addresses on the fly, the mask module filters pruned neuron paths, and the PAU accumulates the corresponding weight contributions into local psums. Together, these modules implement sparse CWP execution and apply the algorithmically generated G-LIF neuron masks in hardware.

\subsubsection{Fetch Module and Streaming Index Decoder}
The fetch module loads CWP-ordered sparse weight blocks and incrementally encoded spike streams from the on-chip buffers. It also detects input boundaries by comparing the feature-map column pointer with the corresponding counter. 
The weight-index decoder reconstructs the valid input-column indices of each resident weight block. Since each block spans \(P_{co}\) output channels, a decoded weight index is shared across these channels, amortizing the metadata-decoding overhead. 
Following the full-delta decoding rule in Eq.~\eqref{eq_decode}, the spike-index decoder reconstructs absolute row indices and identifies EXT codewords. Its register accumulates the decoded index; an EXT codeword clears the valid bit, whereas a non-EXT codeword asserts it to indicate a completed spike index.
The decoder thus reconstructs an ordered spike stream on chip without requiring wider absolute indices to be transferred from off-chip memory.

\subsubsection{Aligned Sparse Event Launch Module}
When decoded spikes are forwarded individually, the downstream pipeline receives fine-grained and potentially unaligned sparse events. Since spikes from different input columns may map to the same spatial position, processing them separately incurs repeated index conversion and mask-index matching and gating, resulting in redundant operations.
Accordingly, the launch module aligns the ordered spike streams associated with each resident weight block through an online $K_a$-way merge, where $K_a$ is the number of input columns in the current launch group. At each step, it selects the minimum valid head index $i_{\min}$ and generates a bit vector $B_J$ indicating which streams contain a spike at that index:
\begin{equation}
B_J(j)=
\begin{cases}
1, & v_j=1 \ \text{and}\ \hat{i}_j=i_{\min},\\
0, & \text{otherwise},
\end{cases}
\end{equation}
where $\hat{i}_j$ and $v_j$ are the head index and valid flag of the $j$-th input stream, respectively. The resulting aligned event is represented as
\begin{equation}
E_{\mathrm{launch}}=(i_{\min},B_J),
\end{equation}
which compactly encodes the shared input-row index and the corresponding active input columns.

Once a launched event is accepted by the downstream pipeline, only the streams selected by $B_J$ advance to their next decoded indices, while the others retain their current head indices. A ready/valid handshake controls this update, allowing the launch module to stall under downstream backpressure without disrupting the ordering of the input streams. The resulting aligned event stream enables subsequent stages to share index conversion and mask-gating operations across the active input columns.

\subsubsection{Sparse Implicit im2col Conversion Module}
Conventional im2col lowers convolution to matrix multiplication by explicitly unfolding overlapping input windows. However, materializing the unfolded matrix duplicates input data and increases on-chip storage, data movement, and memory-bandwidth requirements. To avoid this overhead, the proposed sparse implicit im2col module directly maps each launched event from the input feature-map index domain to an output spatial index.

For each launched event $(i_{\min},B_J)$, the module performs
coordinate recovery, validity checking, and output-index generation once for the spatial position specified by $i_{\min}$. The resulting output index is shared across all columns selected by $B_J$, thereby eliminating redundant index mapping within the aggregation group.
For an input feature map of width $W_{\mathrm{in}}$, the coordinates associated with $i_{\min}$ are recovered as:
\begin{equation}
    x_i=i_{\min}\bmod W_{\mathrm{in}},\qquad
    y_i=\left\lfloor\frac{i_{\min}}{W_{\mathrm{in}}}\right\rfloor.
\end{equation}
Let $(d_x,d_y)$ denote the signed offset of the current kernel position relative to the kernel center. Given the convolution stride $s$, a valid mapping must satisfy
\begin{equation}
\begin{aligned}
    &(x_i-d_x)\bmod s=0,\qquad
      (y_i-d_y)\bmod s=0,\\
    &0\leq\frac{x_i-d_x}{s}<W_{\mathrm{out}},\qquad
      0\leq\frac{y_i-d_y}{s}<H_{\mathrm{out}}.
\end{aligned}
\end{equation}
An event that violates any of these conditions is discarded. Otherwise, its flattened output spatial index is:
\begin{equation}
    o=
    \frac{y_i-d_y}{s}W_{\mathrm{out}}+
    \frac{x_i-d_x}{s}.
\end{equation}

The converter then emits:
\begin{equation}
    E_{\mathrm{conv}}=(o,B_J),
\end{equation}
where $o$ identifies the target row in the local psum buffer and $B_J$ specifies the input columns participating in the accumulation. Consequently, the index-conversion overhead scales with the number of aggregated launch events rather than the number of individual spikes.

\subsubsection{Local Psum Update with Sparse Mask-Stream Gating}

Each event generated by the implicit im2col converter is filtered
by the hardened neuron mask before entering the weight-accumulation
and psum-update datapath.

For each output-channel block of $P_{co}$ parallel lanes,
the sparse mask stream stores only the output positions containing
at least one retained neuron. Each mask entry is represented as
\begin{equation}
    E_{mask}=(m_{idx},m),
\end{equation}
where $m_{idx}$ denotes the output spatial index and
$m\in\{0,1\}^{P_{co}}$ is the corresponding
channel mask. The $p$-th lane is retained if $m_p=1$.

The mask entries are ordered by $m_{idx}$ and matched
against the output index $o$ generated by the converter. If
$o<m_{idx}$, the current output position is absent from
the sparse mask stream and the event is discarded. If
$o>m_{idx}$, the mask stream advances until
$m_{idx}\geq o$ or the stream is exhausted. The $p$-th
output-channel lane is then enabled according to
\begin{equation}
    g_p(o)=
    \begin{cases}
        m_p, & o=m_{idx},\\
        0,   & \text{otherwise},
    \end{cases}
\end{equation}
where $g_p(o)$ is the corresponding update-enable signal.

Only lanes with $g_p(o)=1$ participate in weight accumulation. For
each enabled lane, the weight-update unit accumulates the weight
elements selected by $B_J$:
\begin{equation}
    \Delta P_p=
    \sum_{j:\,B_J(j)=1}
    W_p(j,k_{\mathrm{pp}}),
    \qquad g_p(o)=1,
\end{equation}
where $k_{\mathrm{pp}}$ denotes the current kernel pixel position (pp). The
result is then accumulated into the corresponding local partial sum:
\begin{equation}
    P_p(o)\leftarrow P_p(o)+\Delta P_p,
    \qquad g_p(o)=1.
\end{equation}
When $g_p(o)=0$, the lane remains inactive and issues neither
weight accumulation nor psum buffer accesses.

\begin{table}[tbp]
    \caption{\revision{SNN architectures.}}
    \label{tab_snn_models}
    \centering
    \footnotesize
    \renewcommand{\arraystretch}{1.18}
    \setlength{\tabcolsep}{3.5pt}

    \begin{threeparttable}
        \begin{tabularx}{\columnwidth}{
            @{}
            >{\raggedright\arraybackslash}p{1.30cm}
            >{\raggedright\arraybackslash}X
            @{}
        }
            \toprule
            \textbf{Network} & \textbf{Network structure} \\
            \midrule

            SCNN6 &
            $3{\times}32{\times}32$-
            $[(256\mathrm{c}3){\times}2\text{-}\mathrm{p}2]$-
            $[(256\mathrm{c}3){\times}3\text{-}\mathrm{p}2]$-
            16384-2048-10
            \\

            \addlinespace[1pt]
            SCNN9 &
            $3{\times}32{\times}32$-
            $[(256\mathrm{c}3){\times}2\text{-}\mathrm{p}2]$-
            $[(256\mathrm{c}3){\times}4\text{-}\mathrm{p}2]$-
            $[(256\mathrm{c}3){\times}2\text{-}\mathrm{p}2]$-
            4096-2048-100
            \\

            \addlinespace[1pt]
            VGG-11 &
            $3{\times}32{\times}32$-
            $[64\mathrm{c}3\text{-}\mathrm{p}2]$-
            $[128\mathrm{c}3\text{-}256\mathrm{c}3\text{-}\mathrm{p}2]$-
            $[256\mathrm{c}3\text{-}512\mathrm{c}3\text{-}\mathrm{p}2]$-
            $[(512\mathrm{c}3){\times}2]$-
            $\mathrm{AA}$-512-10/100
            \\

            \addlinespace[1pt]
            VGGSNN &
            $2{\times}48{\times}48$-
            $[64\mathrm{c}3\text{-}\mathrm{p}2]$-
            $[128\mathrm{c}3\text{-}256\mathrm{c}3\text{-}\mathrm{p}2]$-
            $[256\mathrm{c}3\text{-}512\mathrm{c}3\text{-}\mathrm{p}2]$-
            $[(512\mathrm{c}3){\times}2\text{-}\mathrm{p}2]$-
            4608-10/11
            \\

            \addlinespace[1pt]
            ResNet18 &
            $3{\times}32{\times}32$-
            $[(64\mathrm{c}3){\times}4\text{-}\mathrm{p}2]$-
            $[64\mathrm{c}3\text{-}(128\mathrm{c}3){\times}3
              \text{-}\mathrm{p}2]$-
            $[128\mathrm{c}3\text{-}(256\mathrm{c}3){\times}3
              \text{-}\mathrm{p}2]$-
            $[256\mathrm{c}3\text{-}(512\mathrm{c}3){\times}3
              \text{-}\mathrm{p}2]$-
            $\mathrm{AA}$-512-10
            \\

            \addlinespace[1pt]
            ResNet34 &
            $3{\times}56{\times}56$-
            $[(64\mathrm{c}3){\times}6\text{-}\mathrm{p}2]$-
            $[64\mathrm{c}3\text{-}(128\mathrm{c}3){\times}7
              \text{-}\mathrm{p}2]$-
            $[128\mathrm{c}3\text{-}(256\mathrm{c}3){\times}11
              \text{-}\mathrm{p}2]$-
            $[256\mathrm{c}3\text{-}(512\mathrm{c}3){\times}5
              \text{-}\mathrm{p}2]$-
            $\mathrm{AA}$-512-1000
            \\

            \bottomrule
        \end{tabularx}
        \begin{tablenotes}[flushleft]
            \scriptsize
            \setlength{\itemsep}{0pt}
            \item[] In the network notation, $n\mathrm{c}3$ denotes a
            convolutional layer with $n$ input channels and a
            $3{\times}3$ kernel; $\mathrm{p}2$ denotes a pooling layer
            with a $2{\times}2$ kernel; and $\mathrm{AA}$ denotes adaptive
            average pooling.
        \end{tablenotes}
    \end{threeparttable}
    \vspace{-10pt}
\end{table}

\begin{figure}[tbp]
    \centering 
    \includegraphics[width=0.9\linewidth]{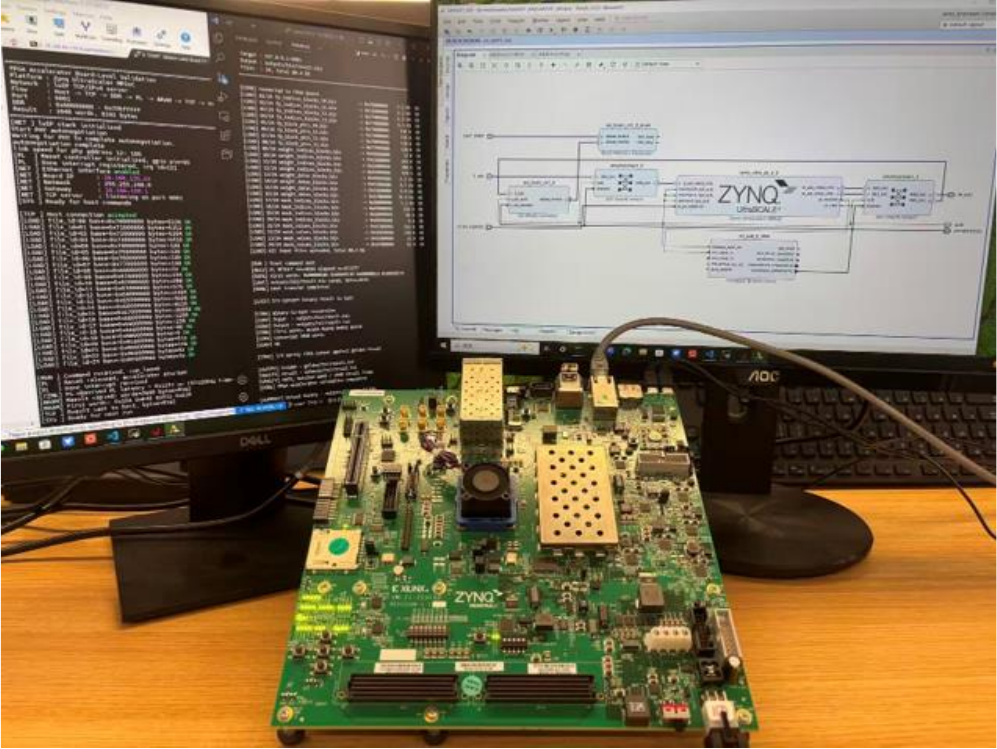}
    \caption{Real-world test setup of a classification system for functional
    validation of the proposed SparseForge.}
    \label{realtest}
    \vspace{-10pt}
\end{figure}

\section{Experimental Evaluation} \label{six}

\subsection{Experimental Setup and Methodology}
We evaluate SparseForge at the algorithm, architecture, and FPGA implementation levels. First, the proposed hardware-aware neuron pruning method is evaluated in terms of accuracy, retained neurons, spike-triggered synaptic operations, and psum buffer saving. Controlled ablation studies are then conducted to evaluate sparse CWP and the spatially and temporally decoupled execution mechanisms. Finally, the complete accelerator is evaluated in terms of inference delay, power consumption, and resource utilization, and compared with representative SNN accelerators.

\revision{
The evaluated SNN architectures are summarized in \tabref{tab_snn_models}, including SCNN6, SCNN9, VGG-11 \cite{vgg}, VGGSNN, ResNet18, and ResNet34 \cite{resnet}. These networks cover different depths, input resolutions, and output dimensions, providing diverse computational and storage workloads for evaluating the proposed pruning and acceleration methods.
}

SparseForge is implemented on a Xilinx ZCU9EG FPGA platform, as shown in \fgref{realtest}. The processing system (PS) configures the accelerator and manages AXI-based data transfers, while the programmable logic (PL) performs sparse SNN inference. Inference latency is measured after the accelerator pipeline is fully occupied by counting the cycles between consecutive completion signals under continuous operation. Dynamic power is estimated using Switching Activity Interchange Format (SAIF)-annotated post-implementation power analysis.

\begin{figure*}[tbp]
    \centering 
    \includegraphics[width=0.9\linewidth]{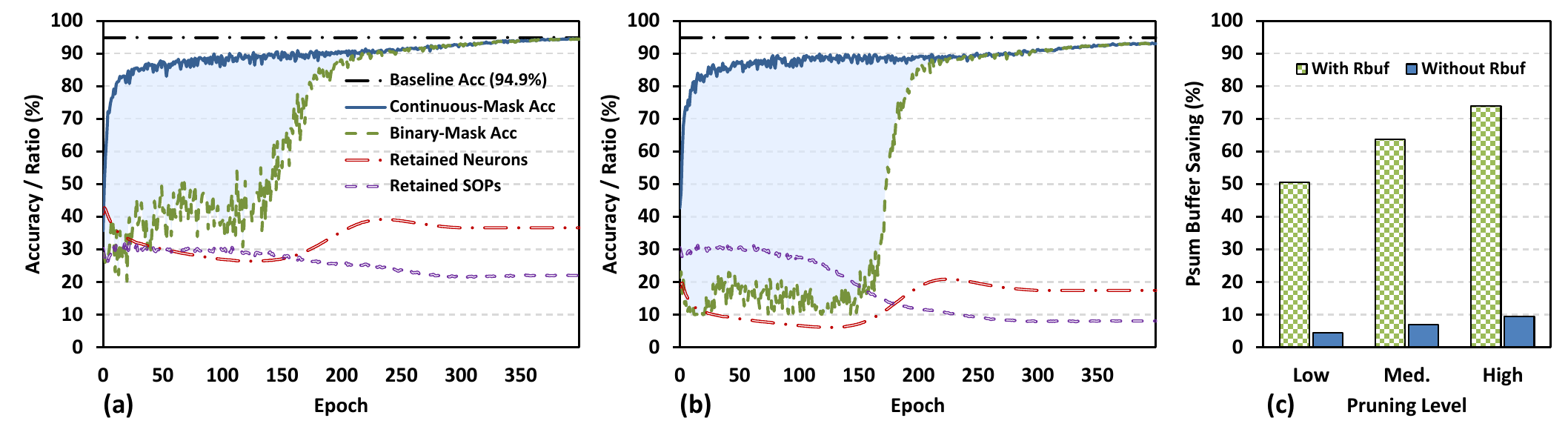}
    \vspace{-8pt}
    \caption{\revision{Hardware-aware pruning results of SCNN6 on CIFAR-10: training evolution under (a) Low and (b) High pruning levels, and (c) psum buffer saving with and without \(R_{\mathrm{buf}}\). The shaded regions indicate the continuous--binary accuracy gaps.}}
    \label{fig_pruning}
    \vspace{-15pt}
\end{figure*}

\subsection{Effectiveness of Hardware-Aware Neuron Pruning}
Unless otherwise specified, the experiments in this subsection are conducted using the SCNN6 architecture on the CIFAR-10 dataset. We evaluate the proposed method from three aspects: the optimization behavior of the G-LIF structural gates, the effectiveness of the buffer-capacity-aware regularizer, and its applicability to different SNN backbones and datasets.

\subsubsection{G-LIF Neuron Pruning}
We first examine how the proposed G-LIF neuron progressively learns deterministic structural pruning decisions during training. 
\revision{
\fgref{fig_pruning} (a) and \fgref{fig_pruning} (b) show the training processes under the Low and High pruning levels, respectively. Each subfigure reports the classification accuracy obtained with the continuous gates and the hardened binary masks, together with the ratios of retained neurons and retained SOPs. The unpruned SCNN6 model, which achieves an accuracy of 94.90\%, is included as the reference.
}

\revision{
Under the Low pruning level in \fgref{fig_pruning} (a), the continuous gates are gradually polarized as the pruning pressure increases. The accuracy initially fluctuates during this transition but subsequently recovers as the network adapts to the learned sparse structure. At convergence, the model retains 36.94\% of the neurons and 21.27\% of the original SOPs, while achieving an accuracy of 94.66\%, corresponding to only a 0.24\% decrease from the unpruned baseline.
}

\revision{
\fgref{fig_pruning} (b) presents the High pruning level, under which stronger regularization produces a more compact network. The final model retains only 17.00\% of the neurons and 7.60\% of the original SOPs, while maintaining an accuracy of 93.28\%. Although the stronger pruning pressure causes a more pronounced continuous--binary accuracy gap during the gate-hardening stage, the gap progressively narrows as training proceeds.
}

\revision{
For both pruning levels, the accuracy obtained with the hardened binary masks eventually approaches that of the continuous-gate model. This observation verifies that the straight-through optimization in Eq.~\eqref{eq:ste_mask} can reliably convert the learned continuous gates into deterministic inference-time masks. Therefore, the proposed G-LIF formulation provides not only a differentiable training mechanism but also an accelerator-compatible structural representation. 
During inference, neurons removed by the hardened masks require no psum accumulation, membrane-potential update, threshold comparison, reset operation, or output-spike propagation.
}

\subsubsection{Buffer-Capacity-Aware Regularization}
As discussed in \secref{3d}, the required psum buffer depth is governed by the maximum retained-neuron density among the output channels sharing the same spatial configuration, rather than by the overall neuron-retention ratio. Accordingly, Eq.~\eqref{eq:smooth_max_density} uses a smooth maximum to approximate this hardware-determining density and incorporate it into the training objective through \(R_{\mathrm{buf}}\).

\revision{
\fgref{fig_pruning} (c) compares the psum buffer savings achieved with and without \(R_{\mathrm{buf}}\) under the Low, Medium, and High pruning levels. When \(R_{\mathrm{buf}}\) is disabled, the pruning objective does not directly constrain the channels with the highest retention densities. These channels therefore continue to determine a relatively large buffer depth, even when many neurons have been removed elsewhere. In contrast, \(R_{\mathrm{buf}}\) explicitly suppresses the maximum per-channel density and increases the psum buffer savings to 50.6\%, 63.8\%, and 74.0\% at the three pruning levels, respectively. This comparison demonstrates that \(R_{\mathrm{buf}}\) is essential for translating neuron pruning into substantial physical psum buffer capacity reduction.
}

\begin{table*}[!t]
    \renewcommand{\arraystretch}{1.18}
    \scriptsize
    \vspace{-5pt}
    \setlength{\tabcolsep}{2.0pt}
    \centering
    \caption{\revision{Cross-network pruning results of SparseForge under different hardware-aware pruning configurations.}}
    \label{tab:algorithm_pruning_results}

    \begin{threeparttable}
    \begin{tabularx}{\linewidth}{l l c c C C c C C c C C}
        \Xhline{1pt}

        Network
        & Dataset
        & \makecell{Time\\Steps}
        & \makecell{Pruning\\Level\tnote{1}}
        & \makecell{Accuracy\\(\%)}
        & \makecell{$\Delta$Acc.\\(pp)\tnote{2}}
        & \makecell{Neurons\\Retained (\%)}
        & $\lambda_{\mathrm{sop}}$
        & \makecell{SOPs\\(M)}
        & \makecell{SOPs\\Retained (\%)}
        & $\lambda_{\mathrm{buf}}$
        & \makecell{Psum Buffer\\Saved (\%)\tnote{3}} \\

        \Xhline{0.5pt}

        \multirow{4}{*}{SCNN6}
        & \multirow{4}{*}{CIFAR-10}
        & 4 & Unpr. & 94.90 &  0.00 & 100.00 & 0 & 1057 & 100.00 & 0 &  0.00 \\
        & & 4 & Low  & 94.66 & -0.24 &  36.94 & $2\!\times\!10^{-2}$ & 225 & 21.27 & $2\!\times\!10^{-4}$ & 50.60 \\
        & & 4 & Med. & 94.03 & -0.87 &  26.88 & $4\!\times\!10^{-2}$ & 146 & 13.80 & $4\!\times\!10^{-4}$ & 63.80 \\
        & & 4 & High & 93.28 & -1.62 &  17.00 & $8\!\times\!10^{-2}$ &  81 &  7.60 & $8\!\times\!10^{-4}$ & 74.00 \\

        \Xhline{0.5pt}

        \multirow{4}{*}{SCNN9}
        & \multirow{4}{*}{CIFAR-100}
        & 4 & Unpr. & 79.29 &  0.00 & 100.00 & 0 & 1151 & 100.00 & 0 &  0.00 \\
        & & 4 & Low  & 79.07 & -0.22 &  59.16 & $1\!\times\!10^{-2}$ & 586 & 50.97 & $4\!\times\!10^{-5}$ &  9.45 \\
        & & 4 & Med. & 78.69 & -0.60 &  46.50 & $4\!\times\!10^{-2}$ & 406 & 35.25 & $1\!\times\!10^{-4}$ & 24.50 \\
        & & 4 & High & 77.40 & -1.89 &  36.54 & $8\!\times\!10^{-2}$ & 287 & 24.99 & $2\!\times\!10^{-4}$ & 44.11 \\

        \Xhline{0.5pt}

        \multirow{4}{*}{VGGSNN}
        & \multirow{4}{*}{\makecell{DVS\\CIFAR-10}}
        & 10 & Unpr. & 83.40 &  0.00 & 100.00 & 0 & 1254 & 100.00 & 0 &  0.00 \\
        & & 10 & Low  & 82.80 & -0.60 &  38.82 & $5\!\times\!10^{-3}$ & 390 & 31.10 & $1\!\times\!10^{-4}$ & 65.75 \\
        & & 10 & Med. & 82.30 & -1.10 &  33.31 & $8\!\times\!10^{-3}$ & 311 & 24.82 & $2\!\times\!10^{-4}$ & 71.70 \\
        & & 10 & High & 82.10 & -1.30 &  27.27 & $2\!\times\!10^{-2}$ & 246 & 19.59 & $3\!\times\!10^{-4}$ & 76.54 \\

        \Xhline{0.5pt}

        \multirow{4}{*}{VGGSNN}
        & \multirow{4}{*}{\makecell{DVS128\\Gesture}}
        & 20 & Unpr. & 96.88 &  0.00 & 100.00 & 0 & 3714 & 100.00 & 0 &  0.00 \\
        & & 20 & Low  & 97.57 & +0.69 &   6.76 & $4\!\times\!10^{-1}$ & 56 & 1.51 & $4\!\times\!10^{-3}$ & 88.60 \\
        & & 20 & Med. & 96.18 & -0.70 &   2.69 & 1 & 12 & 0.32 & $4\!\times\!10^{-2}$ & 95.72 \\
        & & 20 & High & 93.06 & -3.82 &   1.75 & 2 &  6 & 0.16 & $8\!\times\!10^{-2}$ & 96.18 \\

        \Xhline{0.5pt}

        \multirow{4}{*}{VGG-11}
        & \multirow{4}{*}{CIFAR-10}
        & 4 & Unpr. & 95.57 &  0.00 & 100.00 & 0 & 368 & 100.00 & 0 &  0.00 \\
        & & 4 & Low  & 94.75 & -0.82 &  47.96 & $1\!\times\!10^{-2}$ & 110 & 29.85 & $1\!\times\!10^{-4}$ & 35.62 \\
        & & 4 & Med. & 94.47 & -1.10 &  41.21 & $2\!\times\!10^{-2}$ &  88 & 23.86 & $2\!\times\!10^{-4}$ & 43.61 \\
        & & 4 & High & 93.84 & -1.73 &  33.08 & $4\!\times\!10^{-2}$ &  65 & 17.54 & $4\!\times\!10^{-4}$ & 53.31 \\

        \Xhline{0.5pt}

        \multirow{4}{*}{VGG-11}
        & \multirow{4}{*}{CIFAR-100}
        & 4 & Unpr. & 78.36 &  0.00 & 100.00 & 0 & 426 & 100.00 & 0 &  0.00 \\
        & & 4 & Low  & 77.74 & -0.62 &  64.80 & $2\!\times\!10^{-4}$ & 178 & 41.67 & $4\!\times\!10^{-6}$ & 15.21 \\
        & & 4 & Med. & 77.27 & -1.09 &  59.39 & $4\!\times\!10^{-3}$ & 162 & 38.00 & $4\!\times\!10^{-5}$ & 25.60 \\
        & & 4 & High & 76.49 & -1.87 &  43.43 & $4\!\times\!10^{-2}$ & 107 & 25.05 & $4\!\times\!10^{-4}$ & 43.57 \\

        \Xhline{0.5pt}

        \multirow{4}{*}{ResNet18}
        & \multirow{4}{*}{CIFAR-10}
        & 4 & Unpr. & 95.77 &  0.00 & 100.00 & 0 & 353 & 100.00 & 0 &  0.00 \\
        & & 4 & Low  & 94.55 & -1.22 &  69.80 & $2\!\times\!10^{-3}$ & 188 & 53.32 & $2\!\times\!10^{-4}$ & 18.45 \\
        & & 4 & Med. & 93.88 & -1.89 &  59.88 & $4\!\times\!10^{-2}$ & 123 & 34.83 & 0 &  8.53 \\
        & & 4 & High & 93.64 & -2.13 &  52.16 & $2\!\times\!10^{-2}$ & 108 & 30.69 & $4\!\times\!10^{-4}$ & 31.19 \\

        \Xhline{0.5pt}

        \multirow{3}{*}{ResNet34}
        & \multirow{3}{*}{ImageNet}
        & 4 & Unpr. & 65.21 &  0.00 & 100.00 & 0 & 2661 & 100.00 & 0 &  0.00 \\
        & & 4 & Low  & 62.33 & -2.88 &  77.99 & $1\!\times\!10^{-2}$ & 1659 & 62.36 & 0 &  0.78 \\
        & & 4 & High & 62.00 & -3.21 &  73.31 & $8\!\times\!10^{-3}$ & 1579 & 59.34 & $8\!\times\!10^{-5}$ & 16.66 \\

        \Xhline{1pt}
    \end{tabularx}

    \begin{tablenotes}
        \item[1] Low, Med., and High indicate relative pruning levels based on Neurons Retained; Unpr. denotes the unpruned baseline.
        \item[2] $\Delta$Acc. is the accuracy change in percentage points relative to the corresponding unpruned baseline.
        \item[3] Psum Buffer Saved denotes the reduction in the provisioned psum-buffer capacity relative to the corresponding unpruned baseline.
        \item[] $\lambda_{\mathrm{sop}}$ and $\lambda_{\mathrm{buf}}$ weight the SOP cost and psum-buffer capacity cost in the pruning objective, respectively.
    \end{tablenotes}

    \end{threeparttable}
    \vspace{-8pt}
\end{table*}

\subsubsection{\revision{Cross-Network Results}}
\tabref{tab:algorithm_pruning_results} evaluates the proposed method across different network architectures, datasets, and timestep settings. As the pruning strength increases, the retained neurons, SOPs, and psum buffer occupancy generally decrease, with a corresponding accuracy--efficiency trade-off. The SOP reduction is typically greater than the neuron reduction, indicating that $\mathcal{R}_{\mathrm{comp}}$ preferentially removes neurons associated with higher spike-triggered computation rather than uniformly reducing the network size.

The proposed method is particularly effective on event-based datasets, where spatial and temporal redundancy provides greater pruning opportunities. On DVS-CIFAR-10 \cite{dvs_CIFAR10}, Low pruning retains 38.82\% of the neurons and 31.10\% of the SOPs while limiting the accuracy loss to 0.60 percentage points, together with a 65.75\% reduction in psum buffer occupancy. 

The benefit is more pronounced on DVS128-Gesture, where only 6.76\% of the neurons and 1.51\% of the SOPs are retained, while the accuracy improves by 0.69 percentage points and the psum buffer occupancy is reduced by 88.60\%. These results suggest that the time-invariant structural gates can identify neurons that remain redundant across event sequences. Moreover, pruning weakly informative neuron responses can act as a form of regularization, explaining the slight accuracy improvement observed on DVS128-Gesture.

The cross-network results further confirm that the total number of retained neurons alone does not determine psum buffer savings. For ResNet18, Medium pruning retains fewer neurons than Low pruning (59.88\% versus 69.80\%) but achieves a smaller buffer saving (8.53\% versus 18.45\%) when $\lambda_{\mathrm{buf}}=0$. This occurs because the retained neurons may remain concentrated in the output channel that determines the provisioned buffer depth. After $\mathcal{R}_{\mathrm{buf}}$ is enabled at the High pruning level, the buffer saving increases to 31.19\%. Overall, $\mathcal{R}_{\mathrm{comp}}$ and $\mathcal{R}_{\mathrm{buf}}$ address complementary hardware costs, allowing SparseForge to balance accuracy, synaptic computation, and psum buffer occupancy across different workloads.

\begin{figure}[tbp]
    \centering 
    \includegraphics[width=0.9\linewidth]{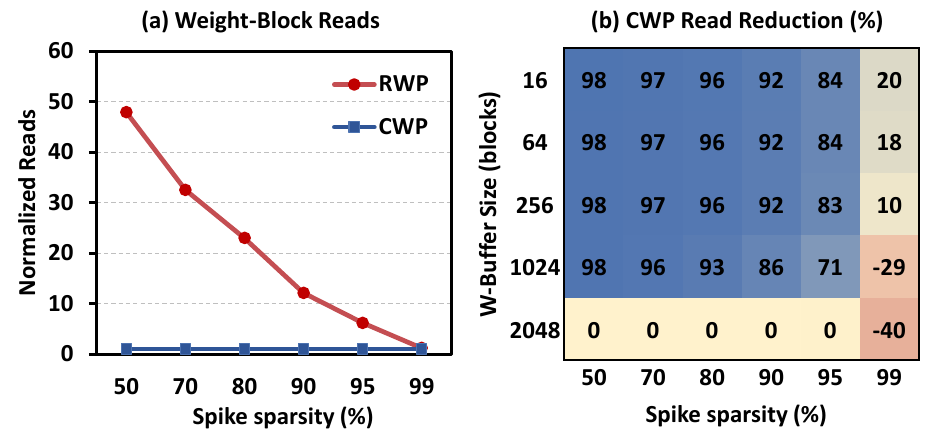}
    \caption{\revision{RWP and CWP comparison: (a) normalized weight-block reads with a 64-block W-Buffer; (b) CWP read reduction across spike sparsity and W-Buffer size.}}
    \label{rwp_cwp_ablation}
    \vspace{-15pt}
\end{figure}

\subsection{\revision{Ablation of Dataflow and Decoupled Execution}}
\label{sec:rwp_cwp_ablation}

\subsubsection{\revision{Sparse CWP versus RWP}}
To isolate the effect of dataflow scheduling, we compare RWP and sparse CWP using randomly generated binary spikes $S\in\{0,1\}^{64\times128}$ and weights $W\in\mathbb{Z}^{128\times64}$ with column-wise $2{:}4$ sparsity along the input-channel dimension. Spike sparsity is swept from 50\% to 99\%, with 60 independent trials per configuration. Both dataflows use identical spike patterns, weight values, sparsity patterns, and W-Buffer capacities, and their outputs are verified to be bit-exact with the reference result $SW$. The compressed weight matrix contains 2,048 blocks. Since the $N{:}M$ metadata and weight values are bound to the same W-Buffer access, CWP reads every block once in physical-address order, whereas RWP generates spike-driven requests through an LRU-managed W-Buffer.

\fgref{rwp_cwp_ablation} (a) reports the number of backing-memory weight-block fetches caused by W-Buffer misses, normalized to the total number of stored weight blocks (2,048). With a 64-block W-Buffer, CWP reduces these fetches relative to RWP by 97.9\% at 50\% spike sparsity and by 18.1\% at 99\% spike sparsity.
\fgref{rwp_cwp_ablation} (b) further shows the read reduction across different W-Buffer capacities, where positive values indicate fewer reads with CWP and negative values indicate fewer reads with RWP. The benefit diminishes as the W-Buffer approaches full weight residency; at 99\% spike sparsity and a 2,048-block W-Buffer, RWP reads 40.2\% fewer blocks because only a subset of the stored weights is activated. Therefore, sparse CWP guarantees deterministic weight traversal, while its block-read benefit depends on the runtime spike density and available W-Buffer capacity.

\begin{figure}[tbp]
    \centering 
    \includegraphics[width=0.95\linewidth]{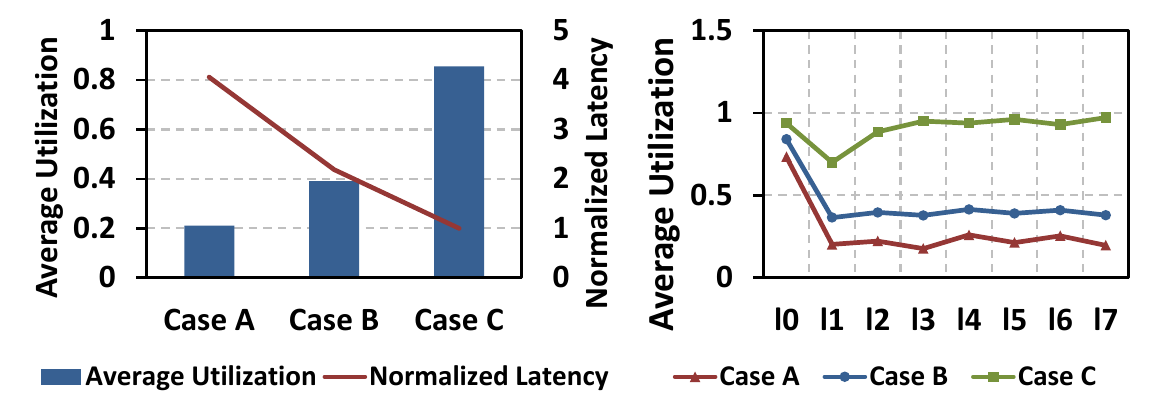}
    \caption{Ablation results of the proposed decoupled execution mechanism on VGGSNN (DVS-CIFAR-10). Case A uses temporally and spatially coupled execution, Case B enables temporal decoupling only, and Case C enables both temporal and spatial decoupling.}
    \label{fig:decoupled_ablation}
    \vspace{-15pt}
\end{figure}

\subsubsection{Temporally and spatially decoupled execution}
To evaluate the effectiveness of the proposed decoupled execution mechanism, we compare three configurations in \fgref{fig:decoupled_ablation}. Case A is the fully coupled baseline, where both timestep contexts and SPUs execute in a lock-step manner. Case B enables temporal decoupling across timestep contexts, while SPUs inside each output-channel group remain spatially coupled. Case C further enables spatial decoupling among SPUs, allowing different input-channel and kernel sub-blocks to accumulate partial sums independently before ordered reduction and readout.

As shown in \fgref{fig:decoupled_ablation} (a), the fully coupled baseline suffers from low average SPU utilization, around 0.21, and the highest normalized latency. This is caused by the highly imbalanced sparse event distribution across timesteps and sub-blocks, where lightly loaded contexts must wait for the slowest ones. With temporal decoupling in Case B, the average utilization increases to about 0.39, and the normalized latency decreases from about 4.0 to 2.1, indicating that timestep-level load imbalance is effectively reduced.

When spatial decoupling is further introduced in Case C, the average SPU utilization reaches about 0.85, and the normalized latency is reduced to 1.0. This improvement shows that spatial lock-step execution among SPUs is still a major bottleneck after temporal decoupling, because different input-channel and kernel sub-blocks may contain different numbers of valid sparse events. Allowing SPUs to progress independently therefore significantly reduces inter-SPU waiting.

\fgref{fig:decoupled_ablation} (b) shows a consistent trend across network layers. Case A exhibits low utilization in most layers, Case B improves utilization by removing timestep-level lock-step constraints, and Case C maintains high utilization across most layers. Overall, the proposed temporally and spatially decoupled execution increases average SPU utilization from about 0.21 to 0.85 and reduces normalized latency to approximately 25\% of the fully coupled baseline, validating its effectiveness for sparse SNN workloads.

\begin{figure}[tbp]
    \centering 
    \includegraphics[width=\linewidth]{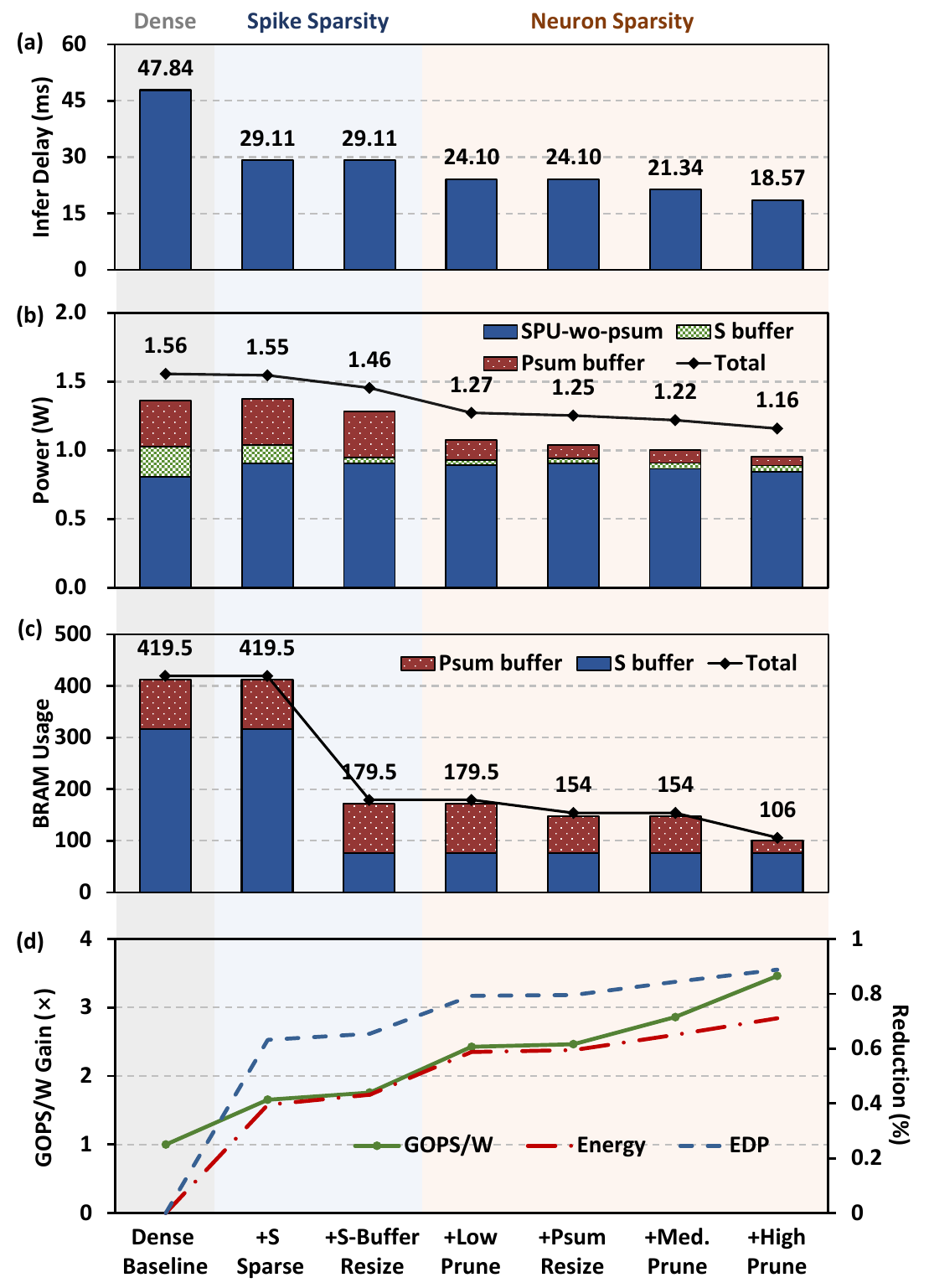}
    \caption{\revision{System-level hardware ablation of SparseForge on SCNN6 with CIFAR-10 under cumulative sparsity optimizations: (a) inference delay, (b) power breakdown, (c) BRAM-usage breakdown, and (d) normalized GOPS/W improvement and reduction ratios of energy per inference and EDP. Diamond markers in (b) and (c) indicate the corresponding total power and BRAM usage.}}
    \label{hardware_ablation}
    \vspace{-15pt}
\end{figure}

\subsection{System-Level Hardware Ablation}
\label{subsec:hardware_ablation}

\revision{
We evaluate seven hardware configurations on SCNN6 with CIFAR-10, as shown along the horizontal axis of \fgref{hardware_ablation}. The configurations are cumulative: each configuration inherits the previously enabled optimizations unless the pruning level is explicitly updated. \textit{Dense Baseline} uses dense spike storage, dense execution, and the unpruned neuron mask. \textit{+S Sparse} enables the proposed sparse spike representation and event-driven execution while retaining the original buffer capacities. \textit{+S-Buffer Resize} further adjusts the physical S-Buffer capacity to the compressed spike representation. \textit{+Low Prune} applies the hardened neuron mask learned at the Low pruning level, while leaving the original psum buffer capacity unchanged. \textit{+Psum Resize} then reduces the psum buffer depth according to the maximum retained-neuron count determined by the Low-pruning mask, as described in \secref{3d}. Finally, \textit{+Med. Prune} and \textit{+High Prune} replace the preceding pruning operating point with the Medium and High levels, respectively, while retaining sparse execution and buffer resizing.
}

\revision{
These configurations progressively expose the hardware benefits of sparse spike execution, compressed buffer provisioning, and hardware-aware neuron pruning. Their effects on inference delay, power, BRAM usage, and energy efficiency are evaluated in \fgref{hardware_ablation} (a)--(d), respectively.
}


\fgref{hardware_ablation} (a) shows the inference delay. Sparse spike execution reduces the delay from 47.84~ms to 29.11~ms by skipping inactive spike events. S-Buffer resizing does not alter the execution workload and therefore introduces no additional delay reduction. Low-level neuron pruning further reduces the delay to 24.10~ms by eliminating computations and state updates associated with pruned neurons, whereas psum buffer resizing preserves the same delay. Increasing the pruning strength to Medium and High reduces the delay to 21.34~ms and 18.57~ms, respectively. Overall, the final configuration achieves a 61.2\% reduction relative to Dense Baseline.

The power breakdown is presented in \fgref{hardware_ablation} (b). Sparse execution only slightly reduces the total power from 1.56~W to 1.55~W because the reduced computation and memory accesses are partially offset by sparse decoding and scheduling overhead. Resizing the S-Buffer reduces its power from 0.132~W to 0.040~W, lowering the total power to 1.46~W. Low-level neuron pruning then reduces runtime switching activity, particularly in psum accumulation, and decreases the total power to 1.27~W. Resizing the psum buffers further reduces their power from 0.144~W to 0.096~W. With Medium- and High-level pruning, the total power reaches 1.22~W and 1.16~W, respectively, corresponding to a 25.6\% reduction from Dense Baseline.

As shown in \fgref{hardware_ablation} (c), sparse execution alone does not reduce BRAM usage because the buffers remain provisioned for the original dense representation. S-Buffer resizing reduces the S-Buffer usage from 316 to 76 BRAMs, lowering the total usage from 419.5 to 179.5. Enabling neuron pruning without resizing the physical psum buffers leaves the BRAM usage unchanged. psum buffer resizing reduces its BRAM usage from 96 to 72 and lowers the total to 154. Under High-level pruning, the psum buffer usage is further reduced to 24 BRAMs, resulting in a total of 106 BRAMs. This represents a 74.7\% reduction relative to Dense Baseline.

\fgref{hardware_ablation} (d) summarizes the cumulative efficiency improvements. All metrics are normalized to Dense Baseline: GOPS/W is reported as an improvement factor, while energy per inference and the energy-delay product (EDP) are reported as reduction ratios. The final High-level pruning configuration improves GOPS/W by \(3.46\times\), while reducing energy per inference and EDP by 71.1\% and 88.8\%, respectively. These results demonstrate that sparse execution primarily reduces inference delay, while buffer resizing converts the learned spike and neuron sparsity into physical storage and power savings. Their combination provides substantial improvements in both performance and energy efficiency.

\begin{table*}[!t]
    \renewcommand{\arraystretch}{1.15}
    \scriptsize
    \vspace{-5pt}
    \setlength{\tabcolsep}{2.2pt}
    \centering
    \caption{\revision{Hardware-level comparison between SparseForge and representative FPGA-based SNN accelerators.}}
    \label{tab:hardware_comparison}

    \begin{threeparttable}
    \resizebox{\linewidth}{!}{
    \begin{tabular}{lccccccccccccccccccc}
        \toprule

        \multicolumn{3}{c}{Implementation}
        & \multicolumn{4}{c}{Benchmark}
        & \multicolumn{3}{c}{Resource}
        & \multicolumn{10}{c}{Performance} \\

        \cmidrule(lr){1-3}
        \cmidrule(lr){4-7}
        \cmidrule(lr){8-10}
        \cmidrule(lr){11-20}

        Work
        & Device
        & \makecell{Freq.\\(MHz)}
        & Network
        & Dataset
        & $T$
        & \makecell{Acc.\tnote{1}\\(\%)}
        & \makecell{LUT\\(K)}
        & BRAM
        & DSP
        & \makecell{Lat.\\(ms)\tnote{2}}
        & \makecell{OPs\\(G)}
        & \makecell{Power\\(W)}
        & FPS
        & FPS/W
        & GOPS
        & GOPS/W
        & \makecell{Energy/\\Infer.\\(mJ)\tnote{2}}
        & \makecell{EDP\\(mJ$\cdot$ms)\tnote{3}}
        & \makecell{GOPS/W/\\kLUT\tnote{4}} \\

        \midrule

        ExSpike \cite{ExSpike}
        & xc7v2000t & 200
        & \multirow{5}{*}{VGG-11}
        & \multirow{9}{*}{CIFAR-10}
        & / & 93.89 & 41 & 312 & 0
        & 6.76 & 2.44 & 1.71 & 148.00 & 86.55
        & 361.51 & 211.91 & 11.56 & 78.15 & 5.17 \\

        SiBrain \cite{SiBrain}
        & virtex-7 & 200
        & & & 4 & 90.25 & 140 & 325 & 1
        & 18.9 & 2.47 & 1.56 & 52.91 & 33.92
        & 130.8 & 83.85 & 29.48 & 557.17 & 0.60 \\

        SConvNSys\cite{SConvNSys}
        & xcu250 & 250
        & & & 4 & 91.48 & 308 & 325 & 0
        & 13.59 & 2.49 & 1.74 & 73.58 & 42.29
        & 183.17 & 105.27 & 23.65 & 326.58 & 0.34 \\

        NEURAL \cite{NEURAL}
        & xc7v2000t & 200
        & & & / & 93.45 & 74 & 138 & 0
        & 14.71 & / & 0.79 & 67.98 & 86.05
        & 41.37 & 52.37 & 11.62 & 170.93 & 0.71 \\

        \textbf{This work}
        & xczu9eg & 190
        & & & 4 & 93.72 & 188 & 76 & 0
        & 4.53 & 2.42 & 1.42 & 220.64 & 155.38
        & 533.95 & 376.02 & 6.43 & 29.13 & 2.00 \\

        \cmidrule(lr){1-4}\cmidrule(lr){6-20}

        ExSpike \cite{ExSpike}
        & xc7v2000t & 200
        & \multirow{3}{*}{ResNet18}
        & & 2 & 94.98 & 41 & 312 & 0
        & 11.76 & 2.46 & 1.73 & 85.00 & 49.13
        & 209.31 & 120.99 & 20.34 & 239.25 & 2.95 \\

        STISA \cite{STISA}
        & xczu9eg & 200
        & & & 2 & 95.29 & 138 & 160 & 0
        & 20.27 & 2.55 & 1.69 & 49.33 & 29.19
        & 126.28 & 75.31 & 34.26 & 694.38 & 0.55 \\

        \textbf{This work}
        & xczu9eg & 200
        & & & 4 & 94.45 & 157 & 186 & 0
        & 9.52 & 4.38 & 1.55 & 105.08 & 67.79
        & 460.25 & 296.94 & 14.76 & 140.52 & 1.89 \\

        \cmidrule(lr){1-4}\cmidrule(lr){6-20}

        \textbf{This work}
        & xczu9eg & 200
        & SCNN6 & & 4 & 93.03 & 95 & 106 & 0
        & 18.57 & 13.28 & 1.16 & 53.85 & 46.42
        & 715.13 & 619.49 & 21.54 & 400.00 & 3.24 \\

        \midrule

        SiBrain \cite{SiBrain}
        & virtex-7 & 200
        & \multirow{3}{*}{VGG-11}
        & \multirow{4}{*}{CIFAR-100}
        & 4 & 66.97 & 142 & 335 & 1
        & 19 & 2.47 & 1.56 & 52.63 & 33.74
        & 130.1 & 83.40 & 29.64 & 563.16 & 0.59 \\

        SConvNSys \cite{SConvNSys}
        & xcu250 & 250
        & & & 4 & 68.54 & 310 & 335 & 0
        & 13.61 & 2.49 & 1.80 & 73.47 & 40.82
        & 182.94 & 101.63 & 24.5 & 333.45 & 0.33 \\

        \textbf{This work}
        & xczu9eg & 190
        & & & 4 & 76.96 & 189 & 92 & 0
        & 5.08 & 2.42 & 1.49 & 196.98 & 132.20
        & 476.69 & 319.93 & 7.57 & 38.46 & 1.69 \\

        \cmidrule(lr){1-4}\cmidrule(lr){6-20}


        \textbf{This work}
        & xczu9eg & 200
        & SCNN9 & & 4 & 77.30 & 184 & 241 & 0
        & 21.74 & 15.10 & 2.12 & 46.0 & 21.70
        & 694.6 & 327.64 & 46.09 & 1002.0 & 1.78 \\

        \midrule

        FireFly2 \cite{Firefly2}
        & xczu5ev & 300
        & \multirow{3}{*}{ResNet34}
        & \multirow{3}{*}{ImageNet}
        & 8 & 67.30 & 26 & 87 & 512
        & 51.02 & 58.72 & 4.04 & 19.60 & 4.85
        & 1150.92 & 284.88 & 206.12 & 10516.28 & 10.96 \\

        STISA \cite{STISA}
        & xczu28dr & 200
        & & & 2.25 & 68.51 & 272 & 230 & 0
        & 77.82 & 18.87 & 2.21 & 12.85 & 5.81
        & 242.48 & 109.72 & 171.98 & 13383.65 & 0.40 \\

        \textbf{This work}
        & xczu9eg & 150
        & & & 4 & 62.13 & 272 & 412 & 0
        & 35.90 & 28.30 & 1.84 & 27.86 & 15.14
        & 788.44 & 428.50 & 66.06 & 2371.55 & 1.58 \\

        \midrule

        \textbf{This work}
        & xczu9eg & 200
        & VGGSNN
        & \makecell{DVS\\CIFAR-10}
        & 10 & 81.50 & 180 & 177 & 0
        & 47.25 & 27.18 & 1.72 & 21.17 & 12.31
        & 575.40 & 334.53 & 81.27 & 3840 & 1.86 \\


        \bottomrule
    \end{tabular}
    }

    \begin{minipage}{\textwidth}
        \footnotesize
        \raggedright
        \textsuperscript{1}
        For the SparseForge entries, the neuron-pruning levels
        from top to bottom are H, L, H, M, H, L, and H,
        respectively, where H, M, and L denote High, Medium,
        and Low. All entries additionally apply 2:4 (50\%)
        weight sparsity. The accuracy differences from the
        corresponding neuron-only results in
        \tabref{tab:algorithm_pruning_results} reflect the
        combined effect of introducing weight sparsity and
        subsequent retraining.
        \textsuperscript{2}
        Lat. denotes inference latency.
        \textsuperscript{3}
        Energy/Infer. denotes energy consumption per inference,
        and EDP denotes the energy--delay product.
        \textsuperscript{4}
        GOPS/W/kLUT denotes resource-normalized energy efficiency.
    \end{minipage}

    \end{threeparttable}
    \vspace{-12pt}
\end{table*}

\subsection{Comparison with State-of-the-Art Accelerators}

\tabref{tab:hardware_comparison} compares SparseForge with representative FPGA-based SNN accelerators from multiple perspectives, including accuracy, resource utilization, throughput, latency, energy efficiency, energy per inference, energy--delay product (EDP), and LUT-normalized energy efficiency. 
\revision{
The operation count of SparseForge follows the convention adopted by the compared works, providing a consistent basis for evaluating GOPS and GOPS/W. Direct quantitative comparisons are primarily made under the same network and dataset settings, while unmatched results are included as complementary references for examining architectural behavior across workloads.
}

\revision{
On VGG-11/CIFAR-10, SparseForge achieves 220.64 FPS, 533.95 GOPS, and 376.02 GOPS/W. Compared with SConvNSys~\cite{SConvNSys}, these results correspond to improvements of $3.00\times$, $2.92\times$, and $3.57\times$, respectively. SparseForge also reduces energy per inference from 23.65 to 6.43 mJ and EDP from 326.58 to 29.13 mJ$\cdot$ms. SConvNSys employs a reconfigurable architecture to accommodate diverse SCNN workloads, whereas SparseForge coordinates hardware-aware neuron pruning with the sparse CWP dataflow. This co-design reduces both synaptic computation and neuron-state accesses instead of relying solely on zero-spike skipping, which may account for the observed improvements. Nevertheless, the broader reconfigurability of SConvNSys represents a different design objective that is not fully reflected by these efficiency metrics.
}

\revision{
Compared with SiBrain~\cite{SiBrain}, SparseForge improves GOPS/W from 83.85 to 376.02 and reduces EDP from 557.17 to 29.13 mJ$\cdot$ms. SiBrain detects inactive spikes and exploits them through spatiotemporal parallel processing. SparseForge additionally adapts its execution workload and buffer capacity to the optimized network structure, which may contribute to its lower latency and energy consumption. SparseForge also provides higher absolute throughput and energy efficiency than NEURAL~\cite{NEURAL}. NEURAL, however, targets elastic hybrid data--event execution and single-timestep inference and requires only 74\,k LUTs. Its smaller resource footprint indicates that NEURAL emphasizes compact and flexible execution, whereas SparseForge trades additional logic resources for higher absolute performance.
}

\revision{
ExSpike~\cite{ExSpike} provides a particularly relevant comparison because it maintains full-event execution and applies adjacent-position event compression to reduce redundant accumulations. On VGG-11/CIFAR-10, SparseForge achieves $1.49\times$ higher FPS and $1.77\times$ higher GOPS/W, while reducing EDP from 78.15 to 29.13 mJ$\cdot$ms. The two designs exploit workload redundancy at different granularities: ExSpike primarily targets irregular event activity, whereas SparseForge reduces the effective workload through its coordinated algorithm and architecture optimizations. In contrast, ExSpike uses only 41\,k LUTs and achieves 5.17 GOPS/W/kLUT, compared with 2.00 for SparseForge. Thus, SparseForge provides higher absolute throughput and energy efficiency in this configuration, while ExSpike remains more efficient when normalized solely by LUT usage.
}

A similar tendency is observed on VGG-11/CIFAR-100. SparseForge reaches 196.98 FPS and 319.93 GOPS/W, compared with 73.47 FPS and 101.63 GOPS/W for SConvNSys. Its EDP is reduced from 333.45 to 38.46 mJ$\cdot$ms, while accuracy increases from 68.54\% to 76.96\%. SparseForge also provides higher accuracy and efficiency than SiBrain on this benchmark. These results indicate that the efficiency improvements observed on CIFAR-10 extend to a more challenging classification task, although differences in FPGA devices and implementation strategies should be considered when interpreting their magnitude.

On ResNet18/CIFAR-10, SparseForge achieves 105.08 FPS, 296.94 GOPS/W, and an EDP of 140.52 mJ$\cdot$ms. Compared with ExSpike under the same network and dataset, SparseForge provides $1.24\times$ higher FPS and $2.45\times$ higher GOPS/W while reducing EDP by $1.70\times$. These improvements indicate that the proposed approach can complement event-driven execution by further reducing the effective workload. However, ExSpike uses considerably fewer LUTs and achieves higher LUT-normalized efficiency through its compact full-event execution and event-compression mechanisms. The comparison therefore reveals different design trade-offs between absolute efficiency and resource-normalized efficiency.

Compared with STISA~\cite{STISA}, our previous dense accelerator, SparseForge improves FPS and GOPS/W by $2.13\times$ and $3.94\times$ on ResNet18/CIFAR-10, respectively, while reducing EDP by $4.94\times$. On the larger ResNet34/ImageNet \cite{imagenet} workload, SparseForge similarly achieves $2.17\times$ higher FPS and $3.91\times$ higher GOPS/W, while reducing energy per inference from 171.98 to 66.06 mJ. These improvements reflect the transition from dense spatiotemporal processing to sparsity-aware execution and suggest that the benefits are maintained as network depth and workload increase.

On ResNet34/ImageNet, SparseForge achieves 27.86 FPS and 428.50 GOPS/W at 150 MHz, outperforming FireFly2~\cite{Firefly2} by $1.42\times$ and $1.50\times$, respectively, without using DSPs. FireFly2 operates at 300 MHz and employs 512 DSPs to achieve higher raw GOPS and GOPS/W/kLUT through high-frequency systolic computation. In comparison, SparseForge obtains its efficiency by avoiding unnecessary operations rather than increasing arithmetic resources. Its lower accuracy mainly reflects the adopted baseline model, which achieves 65.21\% before sparsification, as reported in \tabref{tab:algorithm_pruning_results}; the additional degradation introduced by the subsequent optimization is comparatively limited.

For the remaining configurations without directly matched baselines, SparseForge achieves 619.49 GOPS/W on SCNN6/CIFAR-10, 327.64 GOPS/W on SCNN9/CIFAR-100, and 334.53 GOPS/W on VGGSNN/DVS-CIFAR-10. Rather than serving as strict state-of-the-art comparisons, these results demonstrate that the proposed dataflow supports networks of different depths and both frame-based and event-based inputs. The VGGSNN results further suggest that the architecture is not restricted to the activity patterns generated by static-image encoding, although its efficiency still depends on network structure, timestep count, and optimization level.

Overall, SparseForge provides a DSP-free design that coordinates model optimization with sparse execution and adaptive buffering. The results suggest that aligning the optimized workload with the dataflow and memory organization can translate algorithm-level reductions into hardware performance and energy benefits.

\section{Conclusion} \label{seven}  
This paper presented SparseForge, an algorithm--dataflow--architecture co-design framework that integrates hardware-aware G-LIF neuron pruning, sparse CWP dataflow, and a mask-aware accelerator. By aligning network optimization with execution and buffer organization, SparseForge converts structural redundancy into reductions in computation, state processing, and on-chip storage across both frame- and event-based workloads. The complete system reduces latency, energy per inference, and EDP by 61.2\%, 71.1\%, and 88.8\%, respectively, relative to the dense baseline. FPGA evaluation further demonstrates competitive accuracy and improved throughput and energy efficiency over representative accelerators under matched benchmarks, achieving up to 619.49 GOPS/W without DSP utilization.

\begin{IEEEbiography}[{\includegraphics[width=1in,height=1.25in,clip,keepaspectratio]{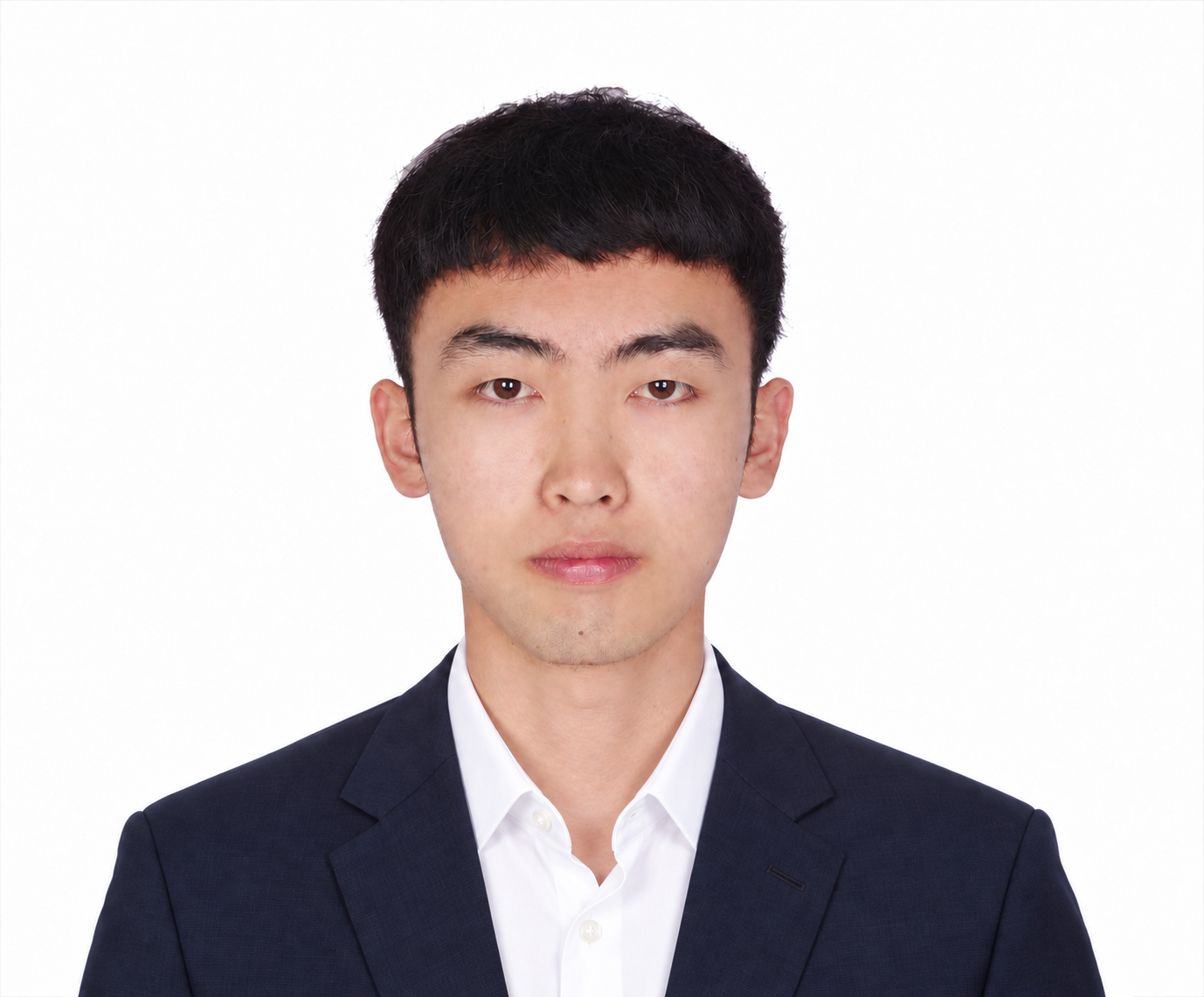}}]{Kainan Wang}
received the B.S. degree in Electronic and Information Engineering from the School of Information Science, Yunnan University, Kunming, China, in 2022. He is currently pursuing the Ph.D. degree in electronics science and technology with the College of Information Science and Electronic Engineering, Zhejiang University, Hangzhou, China. His current research interests include hardware-algorithm co-design for neuromorphic computing and domain-specific processor architectures for edge applications.
\end{IEEEbiography}

\vspace{-10mm} 

\begin{IEEEbiography}[{\includegraphics[width=1in,height=1.25in,clip,keepaspectratio]{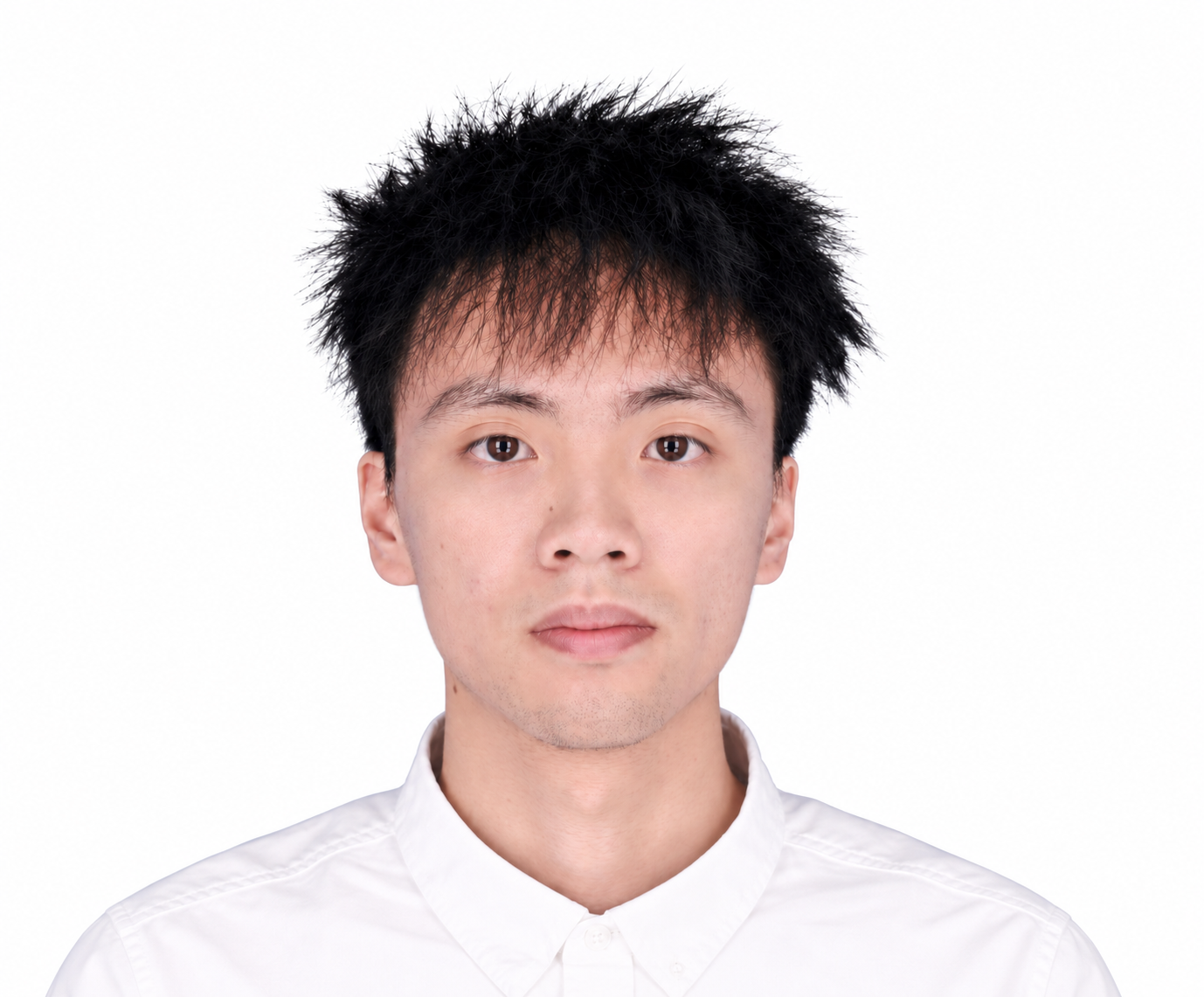}}]{Sanhe Fu}
received the B.S. degree in Electrical Engineering from Zhejiang University, Haining, China, in 2025. He is currently pursuing the M.S. degree in Electronic Information Engineering with Zhejiang University-University of Illinois Urbana-Champaign Institute, Zhejiang University, Haining, China. His current research interests include neural network accelerator design, spiking neural network, brain-like chip.
\end{IEEEbiography}

\vspace{-10mm} 

\begin{IEEEbiography}[{\includegraphics[width=1in,height=1.25in,clip,keepaspectratio]{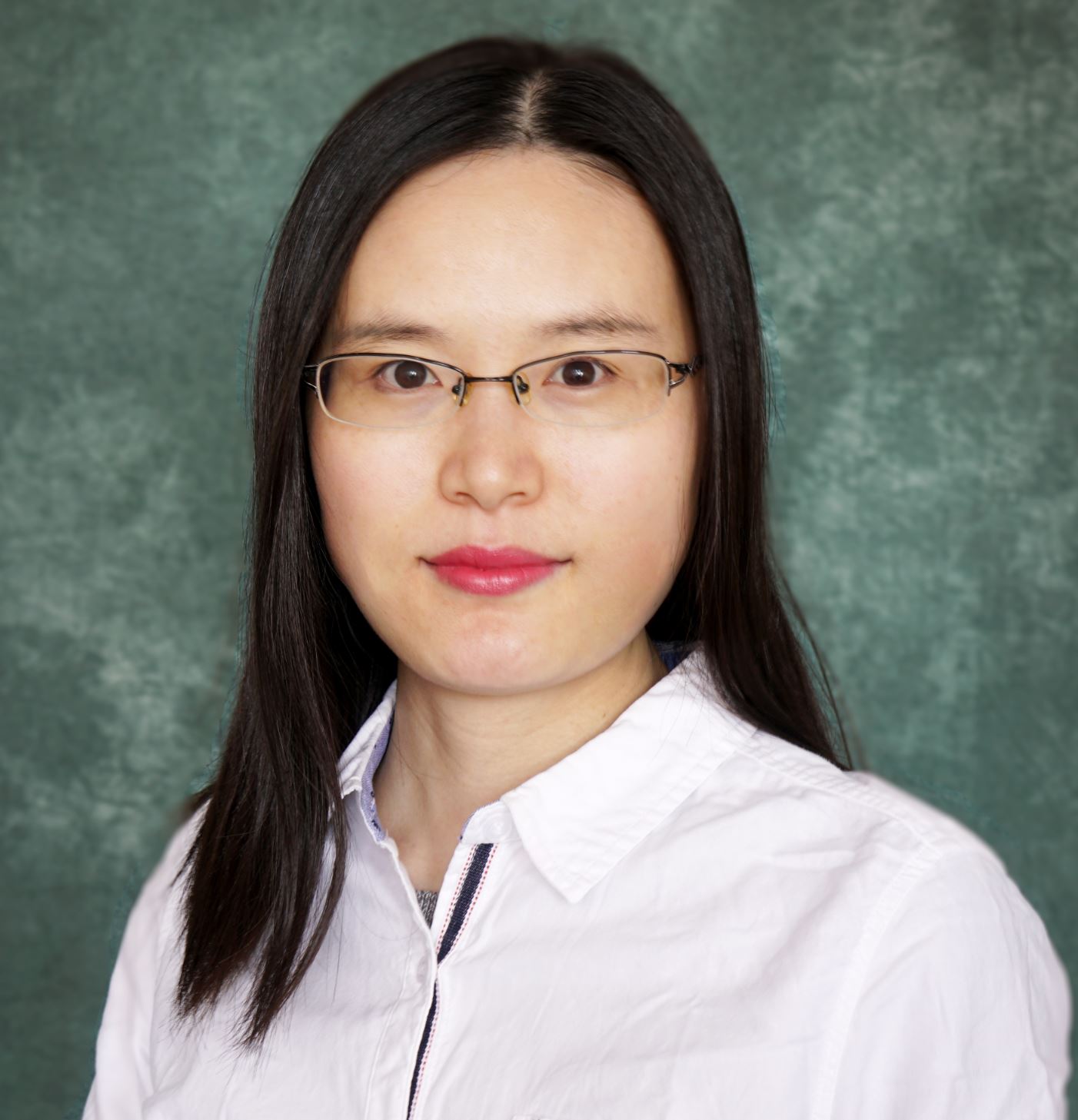}}]{Aili Wang}
(Member, IEEE) received the Ph.D. degree in electrical and computer engineering from the University of Washington, Seattle, WA, USA, in 2019. Since 2020, she has been an Assistant Professor with the Zhejiang University-University of Illinois Urbana-Champaign Institute, Zhejiang University, Haining, China. Her research interests include low-power analog and mixed-signal integrated circuit design and the co-design of algorithms and hardware for brain-inspired computing. She has authored and co-authored many papers published in prestigious journals and conference proceedings, including TCAS-I, TVLSI, Sensors Journal, TAI, NCE, CICC, ISCAS, NeurIPS, ICML, CVPR, and ECAI. Dr. Wang has served as a TPC member of CSTIC since 2022 and of ICICM since 2023. She received the IEEE Solid-State Circuits Society Student Travel Award in 2018.
\end{IEEEbiography} 

\end{document}